\documentclass{article}
\usepackage{spconf,amsmath,graphicx}

\usepackage{enumitem}
\setlist{nosep, leftmargin=14pt}

\usepackage{mwe} % to get dummy images

\usepackage{bm}
\usepackage{amsfonts}
\usepackage{multirow}

\usepackage{subcaption}
\title{A Plug-and-Play Approach to Multiparametric Quantitative MRI: \\ Image Reconstruction using Pre-Trained Deep Denoisers}
\name{Ketan Fatania, Carolin M. Pirkl, Marion I. Menzel, Peter Hall and Mohammad Golbabaee \thanks{KF, PH and MG are with the Department of Computer Science at the University of Bath, UK: (KF432@bath.ac.uk). CMP and MIM are with the Department of Computer Science and Department of Physics at the Technical University of Munich, and GE Healthcare, Germany. MIM is also with AImotion Bavaria at the Technische Hochschule Ingolstadt, Germany.}}
\address{}
\begin{document}
\ninept
\maketitle
\begin{abstract}
Current spatiotemporal deep learning approaches to Magnetic Resonance Fingerprinting (MRF) build artefact-removal models customised to a particular k-space subsampling pattern which is used for fast (compressed) acquisition. This may not be useful when the acquisition process is unknown during training of the deep learning model and/or changes during testing time. This paper proposes an iterative deep learning plug-and-play reconstruction approach to MRF which is adaptive to the forward acquisition process. Spatiotemporal image priors are learned by an image denoiser i.e. a Convolutional Neural Network (CNN), trained to remove generic white gaussian noise (not a particular subsampling artefact) from data. This CNN denoiser is then used as a data-driven shrinkage operator within the iterative reconstruction algorithm. This algorithm with the same denoiser model is then tested on two simulated acquisition processes with distinct subsampling patterns. The results show consistent de-aliasing performance against both acquisition schemes and accurate mapping of tissues' quantitative bio-properties. \textbf{Software available: https://github.com/ketanfatania/QMRI-PnP-Recon-POC}
\end{abstract}

\section{Introduction}
\label{sec:intro}
Magnetic Resonance Fingerprinting (MRF)~\cite{ref:ma2013} is an emerging Quantitative MRI technology for simultaneous measurement of multiple quantitative bio-properties (e.g. T1 and T2 relaxation times, and proton density (PD)) of tissues in a single and time-efficient scan. However, due to the aggressive spatiotemporal subsampling needed for short scan times, the MRF time-series data and consequently the tissue maps usually contain aliasing artefacts. 

Compressed sensing reconstruction algorithms based on using analytical image priors (e.g., sparsity, Total Variation and/or low-rank) have been proposed to address this problem \cite{ref:mcgivney2014, ref:asslander2018, ref:golbabaee2021-lrtv_mrfresnet}. Recent works e.g.~\cite{ref:fang2019oct, ref:chen2020} have been focussed around deep learning approaches that utilise spatiotemporal image priors learned from data for artefact removal. These approaches which are conventionally trained end-to-end on pairs of subsampled and clean data, outperform those using analytical image priors and produce excellent results. However, unlike the traditional compressed sensing algorithms, current trained deep learning models are specific to the subsampling processes used during their training and unable to generalise and remove aliasing artefacts from other subsampling processes given at the testing time.

The contribution of this work is to propose the first deep Plug-and-Play (PnP) iterative reconstruction algorithm for MRF to address this issue, and to demonstrate that this approach effectively adapts to changing acquisition models, specifically, the MRF k-space subsampling patterns. Iterations of the PnP algorithm~\cite{ref:venkatakrishnan2013, ref:ahmad2020} follow steps of the Alternating Direction Method of Multipliers (ADMM)~\cite{ref:boyd2011} which is an optimisation method for model-based compressed sensing reconstruction (for imaging applications of deep learning ADMM and competing methods, see~\cite{ref:ahmad2020, ref:liang2020, ref:romano2017, ref:sun2019, ref:sun2021}). In our work, the spatiotemporal MRF image priors are learned from data through an image denoiser i.e. a Convolutional Neural Network (CNN), that is pre-trained for removing Additive White Gaussian Noise (AWGN), and not any particular subsampling artefact. This CNN denoiser is used as a data-driven shrinkage operator within the ADMM's iterative reconstruction process. The reconstructed (de-aliased) image time-series are then fed to an MRF dictionary matching step~\cite{ref:ma2013} for mapping tissues' quantitative parameters.

\section{The MRF Inverse imaging Problem}
\label{sec:mrf}

MRF adopts a linear spatiotemporal compressive acquisition model:
\begin{equation}
	\label{eq:standard_linear_inverse_problem}
	\bm{y} = \bm{A}\bm{x} + \bm{w}
\end{equation}
where $ \bm{y} \in \mathbb{C}^{\,m\times T}$ are the k-space measurements collected at $T$ temporal frames and corrupted by some bounded noise $\bm{w}$. The acquisition process i.e. the linear forward operator $ \bm{A}: \mathbb{C}^{\,n\times t}\rightarrow \mathbb{C}^{\,m\times T}$ models Fourier transformations subsampled according to a set of temporally-varying k-space locations in each timeframe combined with a temporal-domain compression scheme~\cite{ref:mcgivney2014,ref:asslander2018,ref:golbabaee2021-lrtv_mrfresnet} for (PCA subspace) dimensionality reduction i.e. $t\ll T$. $ \bm{x}\in \mathbb{C}^{\,n\times t} $ is the Time-Series of Magnetisation Images (TSMI) across $n$ voxels and $t$ dimension-reduced timeframes (channels). Accelerated MRF acquisition implies working with under-sampled data which makes the inversion of~\eqref{eq:standard_linear_inverse_problem} i.e. estimating the TSMIs $\bm{x}$ from the compressed MRF measurements $\bm{y}$, an ill-posed problem.  
\vspace{\baselineskip}
\\
\textbf{The Bloch response model:} 
The TSMIs magnetisation time-responses are the main source of tissue quantification.  At each voxel $v$, the time responses i.e. the fingerprints, are related to the corresponding voxel's bio-properties, namely, the T1 and  T2 relaxation times, through the solutions of the Bloch differential equations $\mathcal{B}$ scaled by the proton density (PD)~\cite{ref:ma2013, jiang2015mr}:
\begin{equation}
	\label{eq:magnetisation_response}
	\bm{x}_{v} \approx \text{PD}_{v} \, \mathcal{B} \! \left( \text{T1}_{v}, \text{T2}_v \right)
\end{equation}
While this relationship could temporally constrain the inverse problem~\eqref{eq:standard_linear_inverse_problem}, it turns out to be inadequate to make the problem well-posed~\cite{ref:golbabaee2021-lrtv_mrfresnet}. The model~\eqref{eq:magnetisation_response} alone does not capture cross-voxel correlations. Spatial-domain image priors that account for this must be further added to resolve the ill-posedness problem. For this we rely on datasets of anatomical quantitative maps i.e. the T1, T2 and PD maps, to create the TSMIs via \eqref{eq:magnetisation_response}, and train a denoiser model on them to learn the spatiotemporal structures/priors for $\bm{x}$. We then use this trained denoiser to iteratively solve~\eqref{eq:standard_linear_inverse_problem} for any given forward model $\bm{A}$. This process is detailed in the next section.

\section{image reconstruction Algorithm}
\label{sec:algorithms}

We describe our algorithm to reconstruct artefact-free TMSIs before feeding them to MRF dictionary-matching for quantitative mapping.

\subsection{The PnP-ADMM algorithm}
\label{ssec:pnp_admm}

A model-based reconstruction approach to solve inverse problems like (1) would typically lead to an optimisation of the form
\begin{equation}
	\label{eq:opt}
	\arg\min_{x} \| \bm{y} - \bm{A}\bm{x}\|_2^2 + \phi(\bm{x})
\end{equation}
which can be solved by variety of iterative shrinkage algorithms including ADMM~\cite{ref:boyd2011} and by iterating between a step to minimise the first term and promote the k-space data consistency according to the tested acquisition process, and another shrinkage step according to a regularisation term $\phi$ to promote certain structural priors on $\bm{x}$ to combat the ill-posedness of (1). In the Plug-and-Play (PnP) approach~\cite{ref:venkatakrishnan2013}, the shrinkage term is an AWGN image denoiser $\bm{f}$ that builds an implicit regularisation for (1). The denoiser model $\bm{f}$ could be a trained convolutional neural network (CNN) like~\cite{ref:ahmad2020} that captures the structural image priors for $\bm{x}$ by removing generic gaussian noise from $\bm{x}+\bm{n}$, where $\bm{n} \sim\mathcal{N}(0,\sigma)$, and the noise power $\sigma$ is an experimentally-chosen hyperparameter. 

We use the following ADMM based iterations of the PnP algorithm~\cite{ref:ahmad2020}: $\bm{v}_0 =\bm{A}^Hy$,  $\bm{u}_0=\mathbf{0}$, $\forall k, \text{iteration number} =1,2,...\!\!\!$    % Used \!\!\! to fix extra line produced in arXiv version issue
\begin{subequations}
	\begin{align}
	\label{eq:pnp_admm_a}
	\bm{x}_k & = \bm{h} \left( \bm{v}_{k-1} - \bm{u}_{k-1} \right) \\
	\label{eq:pnp_admm_b}
	\bm{v}_k & = \bm{f} \left( \bm{x}_{k} + \bm{u}_{k-1} \right) \\
	\label{eq:pnp_admm_c}
	\bm{u}_k & = \bm{u}_{k-1} + \left( \bm{x}_{k} - \bm{v}_{k} \right)
	\end{align}
\end{subequations}
where
\begin{equation}
	\label{eq:h}
	\bm{h} \left( \bm{z} \right) = \arg\min_{x}  \|\bm{y}-\bm{Ax}\|_2^2
	 + \gamma \|\bm{x}-\bm{z}\|_2^2 
\end{equation}
The step \eqref{eq:pnp_admm_a} enforces the k-space data consistency, where $ \bm{h} $ is solved using the conjugate gradient algorithm. Here the ADMM's internal convergence parameter is set to $\gamma=0.05$ and $ \bm{z} = \bm{v}_{k-1} - \bm{u}_{k-1} $.  The step \eqref{eq:pnp_admm_b} applies the image denoising shrinkage to promote spatiotemporal TSMI priors, and the final step aggregates the two previous steps to update the next iteration.

\subsection{CNN denoiser}
\label{ssec:cnn_denoiser}
Our PnP algorithm is combined with a pre-trained CNN denoiser which is plugged in as $ \bm{f} $ in \eqref{eq:pnp_admm_b} to iteratively restore the TMSI using learned image priors. The denoiser has a U-Net shape architecture following that of~\cite{ref:zhang2021}. Here we modified the network's input and output dimensions to match the number of TSMI's multiple channels (i.e. we used real-valued TSMIs with $t=10$ in our experiments) to enable multichannel spatiotemporal image processing. A noise level map, filled with $\sigma$ values, of the same dimensions as other channels was appended to the network's input for multi-noise level denoising, following ~\cite{ref:zhang2021}. Other hidden layers follow exactly the same specifications as~\cite{ref:zhang2021}. We trained this model using (multichannel) image patches extracted from $\{(\bm{x},\bm{x}+\bm{n})\}$  pairs of clean and noise-contaminated TSMIs for various levels $\sigma$ of AWGN noise. Further, image patches are patch-wise normalised to the [0,1] range. The clean TSMIs were obtained from a dataset of anatomical quantitative maps via (2).

\section{Numerical Experiments}
\label{sec:experiments}

\textbf{Dataset:} A dataset of quantitative T1, T2 and PD tissue maps of 2D axial brain scans of 8 healthy volunteers across 15 slices each were used in this study\footnote{Data was obtained from a 3T GE scanner (MR750w system - GE Healthcare, Waukesha, WI) with 8-channel receive-only head RF coil, $230\times230$ mm$^2$ FOV, $230\times230$ image pixels, $5$ mm slice thickness, and used a FISP acquisition protocol with $T=1000$ repetitions, the same flip angles as~\cite{jiang2015mr}, inversion time of 18 ms, repetition time of 10 ms and echo time of 1.8 ms. The groundtruth tissue maps were obtained by the LRTV algorithm~\cite{ref:golbabaee2021-lrtv_mrfresnet}. }.
Clean (groundtruth) TSMIs were retrospectively simulated from these maps via~\eqref{eq:magnetisation_response} using the EPG  Bloch model formalism~\cite{weigel2015extended} and a truncated (accelerated) variant of the FISP MRF protocol~\cite{jiang2015mr} with $T=200$ repetitions. PCA was applied to obtain $t=10$ channel dimension-reduced TSMI data~\cite{ref:mcgivney2014}. The TSMIs were real-valued and their spatial resolution was cropped 
from $ 230 \times 230 $ pixels to $ 224 \times 224 $ pixels for the U-Net. The dataset was split into 105 slices (from 7 subjects) for training and 15 slices (for 1 held-out subject) for testing.
\vspace{\baselineskip}
\\
\textbf{Training the CNN denoiser:} The TSMI training data was augmented by random resizing of the spatial resolution. 
The CNN is a multichannel image-patch denoiser. For this we extracted patches of size $128\times128\times10$ with spatial strides of $17$ from the TSMIs. We augmented the patches by flipping across image axes and used random rotations. The patches were then [0,1] normalised. Random AWGN was then added during each iteration of the training algorithm to create pairs of clean and noisy TSMI patches. The CNN was trained following~\cite{ref:zhang2021} to 500 epochs using L1 loss, Adam optimiser, batch size $16$, initialised weights using Kaiming uniform with no bias, and the learning rate initialised at $10^{-4}$ and halved every 100,000 iterations. Randomly selected levels of AWGN noise from $\sigma=10^{\{-4\,\text{to}\,0\}}$ were used for training the denoiser, following~\cite{ref:zhang2021}. For the PnP algorithm, the denoiser was tested with five levels of AWGN noise $\sigma=10^{\{-4,-3,-2,-1,0\}}$ from which $\sigma=10^{-2}$ yielded the best result. A second denoiser trained solely on the optimum noise level found was utilised for our comparisons.
\vspace{\baselineskip}
\\
\textbf{Subsampling models:} We simulated two single-coil Cartesian acquisition processes with distinct k-space subsampling patterns: (i) a spiral readout as in~\cite{ref:chen2020} with rotating patterns across $T=200$ repetitions was used to subsample the $224\times224$ Cartesian FFT grid across all timeframes, and (ii) k-space multiple rows subsampling pattern i.e. the multi-shot Echo Planar Imaging (EPI)~\cite{benjamin2019multi}, with shifted readout lines across the timeframes was used for MRF subsampling. 
Both MRF acquisitions subsampled 771 k-space locations in each timeframe from a total of $224\times224$, leading to a compression ratio of 65, and were contaminated with AWGN noise of $30$ dB SNR.
\vspace{\baselineskip}
%
%% ---------- Results - Spiral Tissue Maps Fig ----------
\begin{figure}[t]

\begin{minipage}[b]{3.5cm}
  \centering
  \centerline{\includegraphics[scale=0.4]{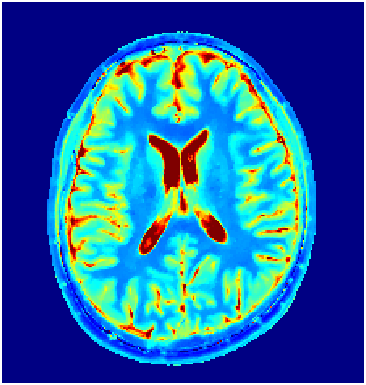}}
%  \vspace{1.5cm}
%  \centerline{(b) Results 3}\medskip
\end{minipage}
\hfill
\begin{minipage}[b]{-3.4cm}
  \centering
  \centerline{\includegraphics[scale=0.4]{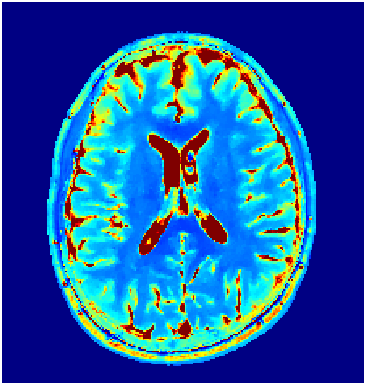}}
%  \vspace{1.5cm}
%. \centerline{(c) Result 4}\medskip
\end{minipage}
\hfill
\begin{minipage}[b]{-3.3cm}
  \centering
  \centerline{\includegraphics[scale=0.4]{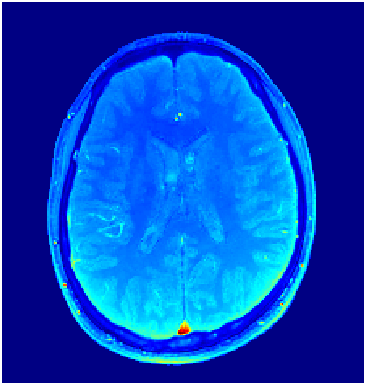}}
%  \vspace{1.5cm}
%. \centerline{(c) Result 4}\medskip
\end{minipage}
\\ % ----------------------------------------------------
\begin{minipage}[b]{3.5cm}
  \centering
  \centerline{\includegraphics[scale=0.4]{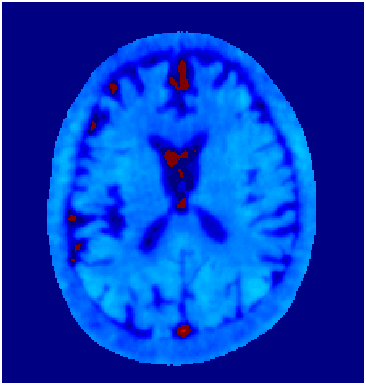}}
%  \vspace{1.5cm}
%  \centerline{(b) Results 3}\medskip
\end{minipage}
\hfill
\begin{minipage}[b]{-3.4cm}
  \centering
  \centerline{\includegraphics[scale=0.4]{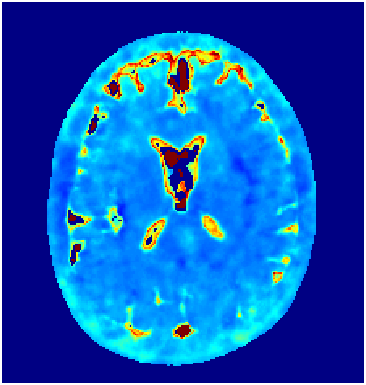}}
%  \vspace{1.5cm}
%. \centerline{(c) Result 4}\medskip
\end{minipage}
\hfill
\begin{minipage}[b]{-3.3cm}
  \centering
  \centerline{\includegraphics[scale=0.4]{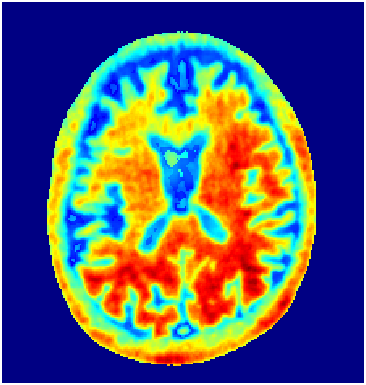}}
%  \vspace{1.5cm}
%. \centerline{(c) Result 4}\medskip
\end{minipage}
\\ % ----------------------------------------------------
\begin{minipage}[b]{3.5cm}
  \centering
  \centerline{\includegraphics[scale=0.4]{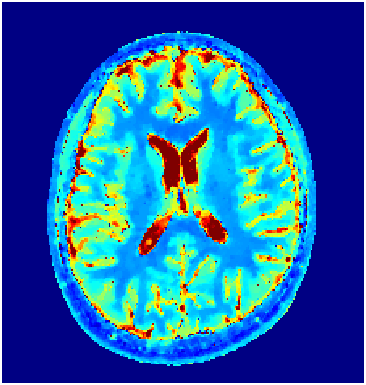}}
%  \vspace{1.5cm}
%  \centerline{(b) Results 3}\medskip
\end{minipage}
\hfill
\begin{minipage}[b]{-3.4cm}
  \centering
  \centerline{\includegraphics[scale=0.4]{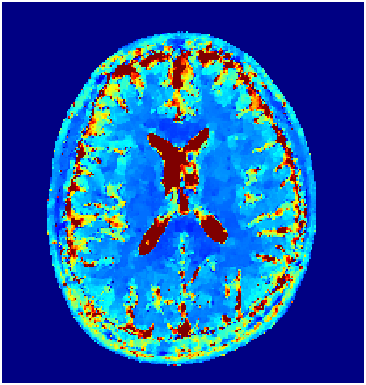}}
%  \vspace{1.5cm}
%. \centerline{(c) Result 4}\medskip
\end{minipage}
\hfill
\begin{minipage}[b]{-3.3cm}
  \centering
  \centerline{\includegraphics[scale=0.4]{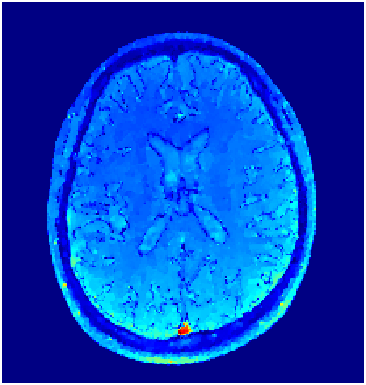}}
%  \vspace{1.5cm}
%. \centerline{(c) Result 4}\medskip
\end{minipage}
\\ % ----------------------------------------------------
\begin{minipage}[b]{3.5cm}
  \centering
  \centerline{\includegraphics[scale=0.4]{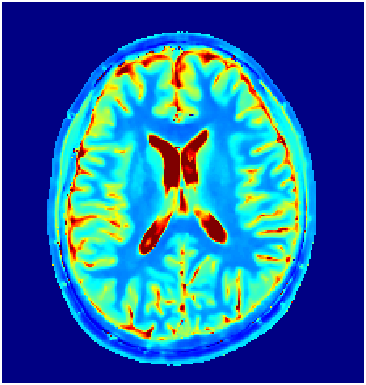}}
%  \vspace{1.5cm}
%  \centerline{(b) Results 3}\medskip
\end{minipage}
\hfill
\begin{minipage}[b]{-3.4cm}
  \centering
  \centerline{\includegraphics[scale=0.4]{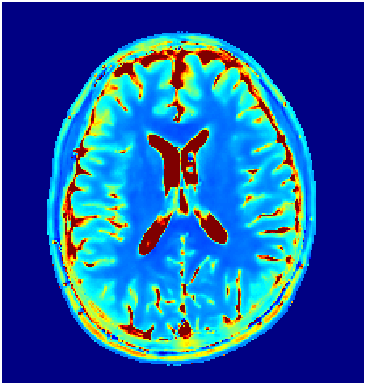}}
%  \vspace{1.5cm}
%. \centerline{(c) Result 4}\medskip
\end{minipage}
\hfill
\begin{minipage}[b]{-3.3cm}
  \centering
  \centerline{\includegraphics[scale=0.4]{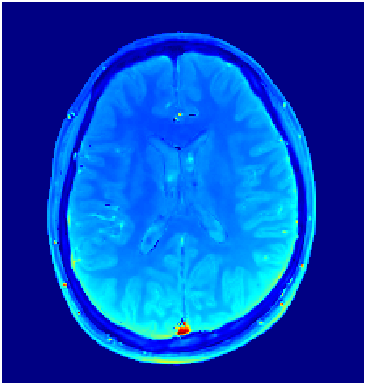}}
%  \vspace{1.5cm}
%. \centerline{(c) Result 4}\medskip
\end{minipage}
\\ % ----------------------------------------------------
\begin{minipage}[b]{3.5cm}
  \centering
  \vspace{1mm}
  \centerline{\includegraphics[scale=0.4]{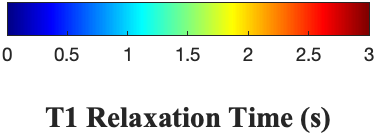}}
%  \vspace{1.5cm}
%  \centerline{(b) Results 3}\medskip
\end{minipage}
\hfill
\begin{minipage}[b]{-3.35cm}
  \centering
  \centerline{\includegraphics[scale=0.4]{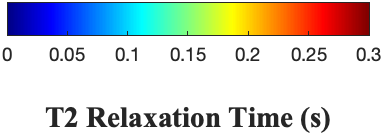}}
%  \vspace{1.5cm}
%. \centerline{(c) Result 4}\medskip
\end{minipage}
\hfill
\begin{minipage}[b]{-3.325cm}
  \centering
  \centerline{\includegraphics[scale=0.4]{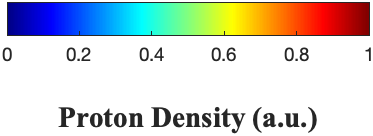}}
%  \vspace{1.5cm}
%. \centerline{(c) Result 4}\medskip
\end{minipage}
\caption{Tissue Map Results using Spiral Subsampling - Col 1-3: T1, T2, PD; Row 1-4: Ground Truth, SVD-MRF, LRTV, PnP-ADMM}
\label{fig:spiral_results}
\vspace{-\baselineskip}
\end{figure}

%% ---------- Results - Spiral Tissue Maps Fig ----------
%%
%% ---------- Results - EPI Tissue Maps Fig ----------
\begin{figure}[t]

\begin{minipage}[b]{3.5cm}
  \centering
  \centerline{\includegraphics[scale=0.4]{./figures/tissue_maps/ground_truth/t1_ref_with_mask_caxis0to3.png}}
%  \vspace{1.5cm}
%  \centerline{(b) Results 3}\medskip
\end{minipage}
\hfill
\begin{minipage}[b]{-3.4cm}
  \centering
  \centerline{\includegraphics[scale=0.4]{./figures/tissue_maps/ground_truth/t2_ref_with_mask_caxis0to0.3.png}}
%  \vspace{1.5cm}
%. \centerline{(c) Result 4}\medskip
\end{minipage}
\hfill
\begin{minipage}[b]{-3.3cm}
  \centering
  \centerline{\includegraphics[scale=0.4]{./figures/tissue_maps/ground_truth/pd_ref_with_mask_caxis0to1.png}}
%  \vspace{1.5cm}
%. \centerline{(c) Result 4}\medskip
\end{minipage}
\\ % ----------------------------------------------------
\begin{minipage}[b]{3.5cm}
  \centering
  \centerline{\includegraphics[scale=0.4]{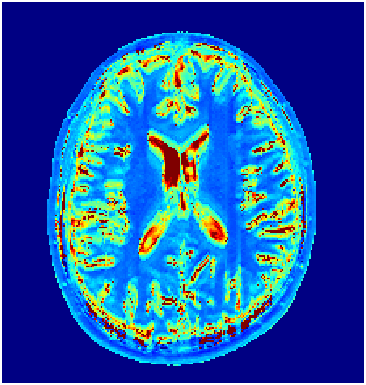}}
%  \vspace{1.5cm}
%  \centerline{(b) Results 3}\medskip
\end{minipage}
\hfill
\begin{minipage}[b]{-3.4cm}
  \centering
  \centerline{\includegraphics[scale=0.4]{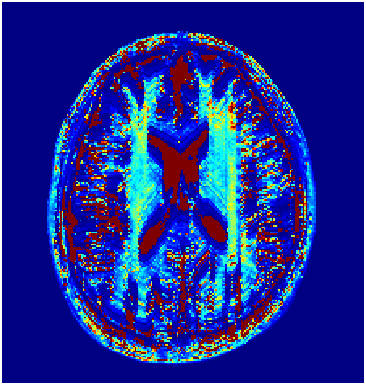}}
%  \vspace{1.5cm}
%. \centerline{(c) Result 4}\medskip
\end{minipage}
\hfill
\begin{minipage}[b]{-3.3cm}
  \centering
  \centerline{\includegraphics[scale=0.4]{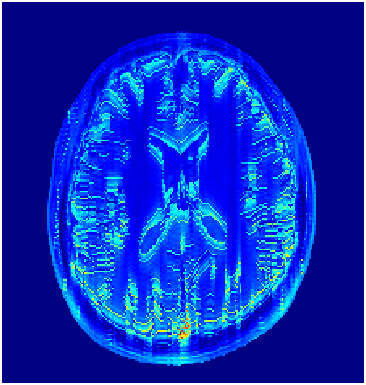}}
%  \vspace{1.5cm}
%. \centerline{(c) Result 4}\medskip
\end{minipage}
\\ % ----------------------------------------------------
\begin{minipage}[b]{3.5cm}
  \centering
  \centerline{\includegraphics[scale=0.4]{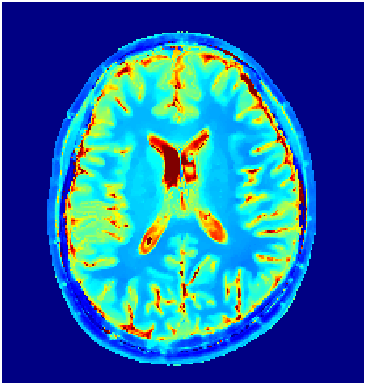}}
%  \vspace{1.5cm}
%  \centerline{(b) Results 3}\medskip
\end{minipage}
\hfill
\begin{minipage}[b]{-3.4cm}
  \centering
  \centerline{\includegraphics[scale=0.4]{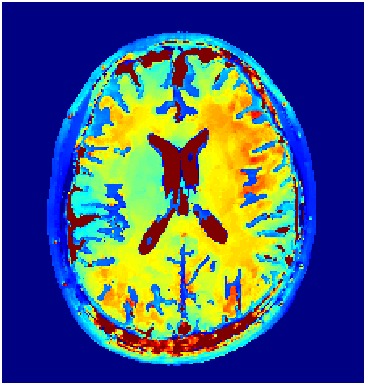}}
%  \vspace{1.5cm}
%. \centerline{(c) Result 4}\medskip
\end{minipage}
\hfill
\begin{minipage}[b]{-3.3cm}
  \centering
  \centerline{\includegraphics[scale=0.4]{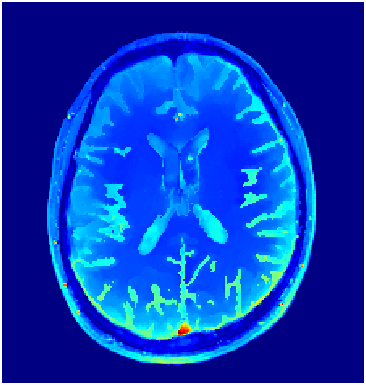}}
%  \vspace{1.5cm}
%. \centerline{(c) Result 4}\medskip
\end{minipage}
\\ % ----------------------------------------------------
\begin{minipage}[b]{3.5cm}
  \centering
  \centerline{\includegraphics[scale=0.4]{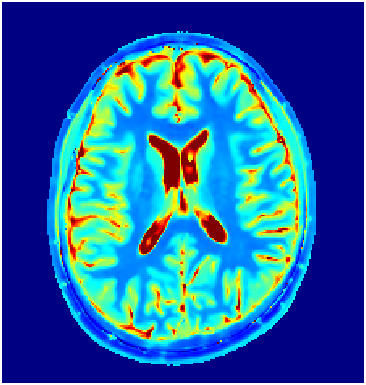}}
%  \vspace{1.5cm}
%  \centerline{(b) Results 3}\medskip
\end{minipage}
\hfill
\begin{minipage}[b]{-3.4cm}
  \centering
  \centerline{\includegraphics[scale=0.4]{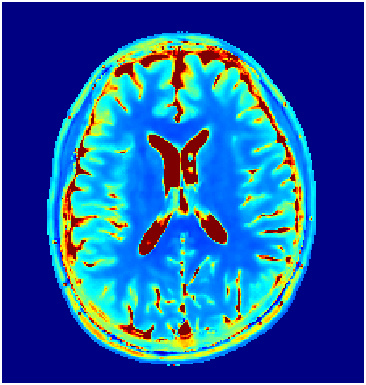}}
%  \vspace{1.5cm}
%. \centerline{(c) Result 4}\medskip
\end{minipage}
\hfill
\begin{minipage}[b]{-3.3cm}
  \centering
  \centerline{\includegraphics[scale=0.4]{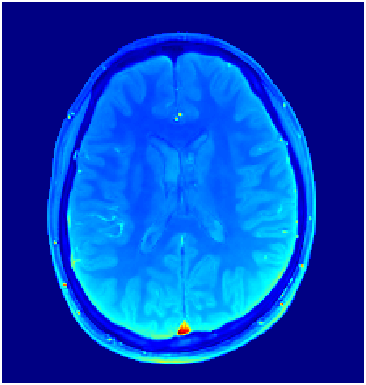}}
%  \vspace{1.5cm}
%. \centerline{(c) Result 4}\medskip
\end{minipage}
\\ % ----------------------------------------------------
\begin{minipage}[b]{3.5cm}
  \centering
  \vspace{1mm}
  \centerline{\includegraphics[scale=0.4]{./figures/tissue_maps/ground_truth/t1_ref_with_mask_caxis0to3_colorbar_horizontal_south.png}}
%  \vspace{1.5cm}
%  \centerline{(b) Results 3}\medskip
\end{minipage}
\hfill
\begin{minipage}[b]{-3.35cm}
  \centering
  \centerline{\includegraphics[scale=0.4]{./figures/tissue_maps/ground_truth/t2_ref_with_mask_caxis0to0.3_colorbar_horizontal_south.png}}
%  \vspace{1.5cm}
%. \centerline{(c) Result 4}\medskip
\end{minipage}
\hfill
\begin{minipage}[b]{-3.325cm}
  \centering
  \centerline{\includegraphics[scale=0.4]{./figures/tissue_maps/ground_truth/pd_ref_with_mask_caxis0to1_colorbar_horizontal_south.png}}
%  \vspace{1.5cm}
%. \centerline{(c) Result 4}\medskip
\end{minipage}
\caption{Tissue Map Results using EPI - Col 1-3: T1, T2, PD; Row 1-4: Ground Truth, SVD-MRF, LRTV, PnP-ADMM}
\label{fig:epi_results}
\vspace{-\baselineskip}
\end{figure}

%% ---------- Results - EPI Tissue Maps Fig ----------
%
\\
\textbf{Tested reconstruction algorithms:} 
We compared the performance of the proposed PnP-ADMM algorithm to the SVD-MRF~\cite{ref:mcgivney2014} and LRTV~\cite{ref:golbabaee2021-lrtv_mrfresnet} TSMI reconstruction baselines. All algorithms incorporated subspace dimensionality reduction ($t=10$). Further they can all adapt to the MRF subsampled acquisition process at testing time. SVD-MRF is a non-iterative approach based on zero-filling reconstruction $\hat{\bm{x}}=\bm{A}^Hy$. The LRTV is an iterative approach to~\eqref{eq:opt} based on an analytical Total Variation image (prior) regularisation for $\phi$. The PnP approach uses data-driven image priors learned by the CNN denoiser. The PnP-ADMM  ran for $100$ iterations, used $\gamma=0.05$ and its conjugate gradient steps ran to the solution tolerance $10^{-4}$. The LRTV ran for $200$ iterations and used regularisation parameter $\lambda=4\times10^{-5}$.
\vspace{\baselineskip}
\\
\textbf{Quantitative mapping:} Dictionary matching was used for mapping the reconstructed TSMIs to the T1, T2 and PD images. For this an MRF dictionary of $94'777$ atoms (fingerprints) was created using the EPG Bloch response simulations for a logarithmically-spaced grid of $($T1, T2$)\in[0.01, 6]\times [0.004,0.6]$ (s). PCA was applied to compress this dictionary's (i.e. the Bloch responses) temporal dimension to a $t=10$ dimensional subspace~\cite{ref:mcgivney2014}. The same subspace was used for the TSMIs dimensionality reduction and within all tested reconstruction algorithms.
\vspace{\baselineskip}
%
%% ---------- Results - Table ----------
\begin{table*}[t] 
	
	\centering
	
	\begin{tabular}{ c c || c c c || c c c } 
		
		\hline \hline \multicolumn{2}{ c ||}{} & \multicolumn{3}{ c ||}{Spiral} & \multicolumn{3}{ c }{EPI} \\ \hline \hline
		
		 & & SVD-MRF & LRTV & PnP-ADMM & SVD-MRF & LRTV & PnP-ADMM \\ \hline \hline
		 
		\multirow{ 2}{*}{MAE (ms)} & T1 & 0.7216 & 0.1235 & \bf{0.0630} & 0.3916 & 0.1756 & \bf{0.0549} \\
		& T2 & 0.1477 & 0.0425 & \bf{0.0240} & 0.3208 & 0.2308 & \bf{0.0250} \\ 
		\hline
		
		\multirow{ 4}{*}{PSNR (dB)} & TSMI & 38.7898 & 50.5719 & \bf{61.9822} & 46.0254 & 40.6593 & \bf{61.2355} \\
		& T1 & 2.3836 & 12.5453 & \bf{17.3983} & 4.8648 & 13.0147 & \bf{19.3664} \\
		& T2 & 9.6561 & 21.7319 & \bf{27.6446} & 6.5304 & 10.3662 & \bf{27.8087} \\
		& PD & 12.5959 & 30.5244 & \bf{36.0676} & 21.5727 & 27.7092 & \bf{40.3840} \\ \hline
		
		\multirow{ 4}{*}{SSIM} & TSMI & 0.6896 & 0.9656 & \bf{0.9959} & 0.8375 & 0.8215 & \bf{0.9964} \\
		& T1 & 0.5695 & 0.8859 & \bf{0.9551} & 0.7588 & 0.9119 & \bf{0.9640} \\
		& T2 & 0.7673 & 0.8410 & \bf{0.9364} & 0.6021 & 0.6859 & \bf{0.9391} \\
		& PD & 0.5978 & 0.8550 & \bf{0.9668} & 0.7001 & 0.8845 & \bf{0.9776} \\ \hline \hline
	
	\end{tabular}
	
	\caption[...]{The metrics for time-series data and tissue maps obtained using Spiral and EPI subsampling patterns averaged over 15 slices.}
	\label{tab:results_table}

\end{table*}
%% ---------- Results - Table ----------
%
\\
\textbf{Evaluation Metrics:} For evaluation of the TSMI reconstruction performance, the Peak Signal-to-Noise-Ratio (PSNR) and Structural Similarity Index Measure (SSIM) averaged across all 10 temporal channels were used. To evaluate tissue maps, the Mean Absolute Error (MAE), PSNR and SSIM were used. Metrics were calculated for all 15 slices of the held-out test subject and averaged.
\vspace{\baselineskip}
\\
\textbf{Results and Discussion:} Fig.\ref{fig:spiral_results}, Fig.\ref{fig:epi_results} and Table.\ref{tab:results_table}, shows PnP-ADMM utilising the same CNN denoiser, trained on generic AWGN, can apply to two different forward models with drastically different subsampling patterns. The output is consistent de-aliasing performance for time-series data and consequently the tissue maps (see supplementary material for a comparison of TSMIs).

PnP-ADMM outperforms tested baselines subjectively (Fig.\ref{fig:spiral_results}, Fig.\ref{fig:epi_results}) and objectively (Table.\ref{tab:results_table}) across all tested metrics, for both tested subsampling patterns. Recovering tissue maps from EPI subsampled data is observed to be generally more challenging than the spiral subsampling scheme (because the centre of k-space is densely sampled by spiral readouts), however, as observed the PnP-ADMM algorithm succeeds while other tested baselines fail. The key to the superior performance of PnP-ADMM lies in its ability to utilise spatiotemporal prior information related to the dataset. SVD-MRF utilises only temporal prior information through the use of PCA, while LRTV utilises generic prior information in the form of temporal priors through PCA and spatial priors through Total Variation. The use of dataset specific spatiotemporal priors, learned by a CNN denoiser, is the crux of PnP-ADMM's superior performance.

\section{Conclusion}
\label{sec:conclusion}

A proof-of-concept is proposed for a PnP-ADMM approach using deep CNN image denoisers for multi-parametric tissue property quantification using MRF compressed sampled acquisitions, which may be useful for cases where the measurement acquisition scheme is unknown during training for deep learning methods and/or may change during testing. 
The method was validated on simulated data and consistently outperformed the tested baselines. This was possible due to the use of data-driven spatiotemporal priors learnt by a pre-trained CNN denoiser, which were critical for enhancing the reconstruction of the TSMIs. Future work will include a variety of measurement acquisition settings, the use of non-gridded sampling trajectories and prospective in-vivo scans.

\section{Compliance with ethical standards}
\label{sec:ethics}

This research study was conducted retrospectively using anonymised human subject scans made available by GE Healthcare who obtained informed consent in compliance with the German Act on Medical Devices. Approval was granted by the Ethics Committee of The University of Bath (Date. Sept 2021 / No. 6568).

\section{Acknowledgments}
\label{sec:acknowledgments}

Carolin M. Pirkl and Marion I. Menzel receive funding from the European Union’s Horizon 2020 research and innovation programme, grant agreement No. 952172.

\bibliographystyle{IEEEbib}
\bibliography{refs}

\end{document}

\ninept
%
% Reference: https://tex.stackexchange.com/questions/53966/how-to-place-a-banner-image-at-the-top-of-a-paper?rq=1
% Reference: https://tex.stackexchange.com/questions/53979/span-columns-with-a-center-environment/53984#53984
% Reference: https://tex.stackexchange.com/questions/55764/input-a-figure-between-title-and-body-in-twocolumn-form
\twocolumn[{%
\renewcommand\twocolumn[1][]{#1}%
\maketitle
%
% ---------------------------------------------------------- Spiral ----------------------------------------------------------
%
% ---------- Supplementary Material - Spiral TSMIs Fig - 1 (Camera Ready - Title Page) ----------
% Reference: https://tex.stackexchange.com/questions/53966/how-to-place-a-banner-image-at-the-top-of-a-paper?rq=1
% Reference: https://tex.stackexchange.com/questions/53979/span-columns-with-a-center-environment/53984#53984
% Reference: https://tex.stackexchange.com/questions/55764/input-a-figure-between-title-and-body-in-twocolumn-form
\makeatletter 									
\newcommand*\captiontype[1]{\def\@captype{#1}} 
\makeatother 	

\begin{center}
	\captiontype{figure}
		\vspace{-3.0mm} 	% -2.0mm (to -3.0mm), -5.5mm, -3.0mm
		%
		% ------------------------------------------------ Ch 1 --------------------------------------------------------
		%
		\begin{minipage}[b]{6.4cm}
			\centering
			\centerline{\includegraphics[scale=0.55]{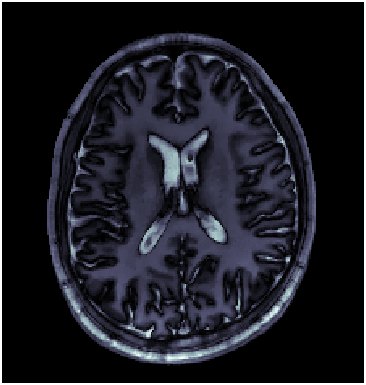}}
			%\vspace{-0.6mm}
			%  \centerline{(b) Results 3}\medskip 		% To add a manual subfigure caption
		\end{minipage}
		\hfill
		\begin{minipage}[b]{-6.4cm}
			\centering
			\centerline{\includegraphics[scale=0.55]{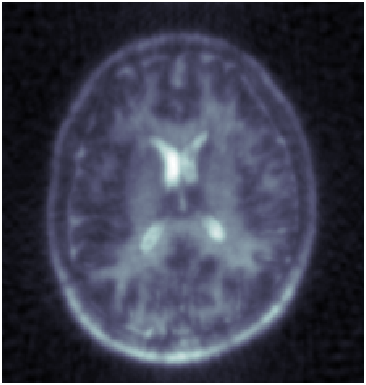}}
			%  \vspace{1.5cm}
			%  \centerline{(b) Results 3}\medskip 		% To add a manual subfigure caption
		\end{minipage}
		\hfill
		\begin{minipage}[b]{-6.4cm}
			\centering
			\centerline{\includegraphics[scale=0.55]{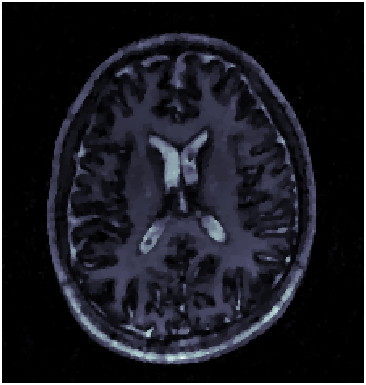}}
			%  \vspace{1.5cm}
			%. \centerline{(c) Result 4}\medskip 			% To add a manual subfigure caption
		\end{minipage}
		\hfill
		\begin{minipage}[b]{-6.4cm}
			\centering
			\centerline{\includegraphics[scale=0.55]{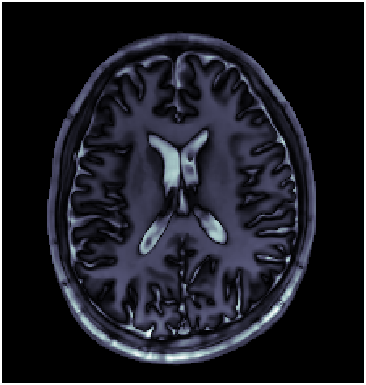}}
			%  \vspace{1.5cm}
			%. \centerline{(c) Result 4}\medskip 			% To add a manual subfigure caption
		\end{minipage}
		\\ % ----------------------------------------------------
		\begin{minipage}[b]{6.4cm}
			\centering
			\vspace{1.0mm}
			\centerline{\includegraphics[scale=0.55]{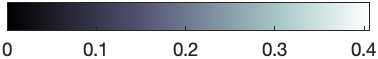}}
			\vspace{2.0mm}
			\centerline{(a) GT: Ch1}\medskip 		% To add a manual subfigure caption
		\end{minipage}
		\hfill
		\begin{minipage}[b]{-6.4cm}
			\centering
			\vspace{1.0mm}
			\centerline{\includegraphics[scale=0.55]{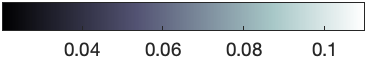}}
			\vspace{2.0mm}
			\centerline{(b) SVD-MRF: Ch1}\medskip 		% To add a manual subfigure caption
		\end{minipage}
		\hfill
		\begin{minipage}[b]{-6.4cm}
			\centering
			\vspace{1.0mm}
			\centerline{\includegraphics[scale=0.55]{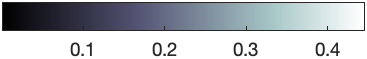}}
			\vspace{2.0mm}
			\centerline{(c) LRTV: Ch1}\medskip 		% To add a manual subfigure caption
		\end{minipage}
		\hfill
		\begin{minipage}[b]{-6.4cm}
			\centering
			\vspace{1.0mm}
			\centerline{\includegraphics[scale=0.55]{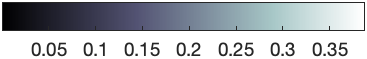}}
			\vspace{2.0mm}
			\centerline{(d) PnP-ADMM: Ch1}\medskip 		% To add a manual subfigure caption
		\end{minipage}
		%
		% ------------------------------------------------ Ch 1 --------------------------------------------------------
		%
		\\
		\vspace{3.5mm} %1.5, 2.5mm
		%
		% ------------------------------------------------ Ch 2 --------------------------------------------------------
		%
		\begin{minipage}[b]{6.4cm}
			\centering
			\centerline{\includegraphics[scale=0.55]{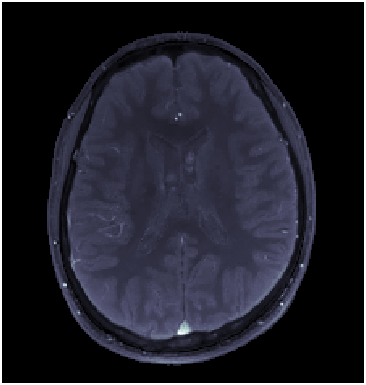}}
			%\vspace{-0.6mm}
			%  \centerline{(b) Results 3}\medskip 		% To add a manual subfigure caption
		\end{minipage}
		\hfill
		\begin{minipage}[b]{-6.4cm}
			\centering
			\centerline{\includegraphics[scale=0.55]{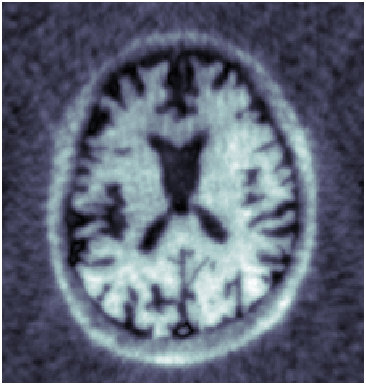}}
			%  \vspace{1.5cm}
			%  \centerline{(b) Results 3}\medskip 		% To add a manual subfigure caption
		\end{minipage}
		\hfill
		\begin{minipage}[b]{-6.4cm}
			\centering
			\centerline{\includegraphics[scale=0.55]{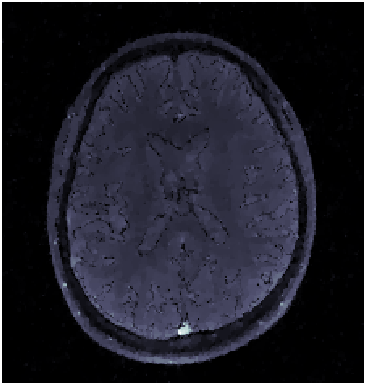}}
			%  \vspace{1.5cm}
			%. \centerline{(c) Result 4}\medskip 			% To add a manual subfigure caption
		\end{minipage}
		\hfill
		\begin{minipage}[b]{-6.4cm}
			\centering
			\centerline{\includegraphics[scale=0.55]{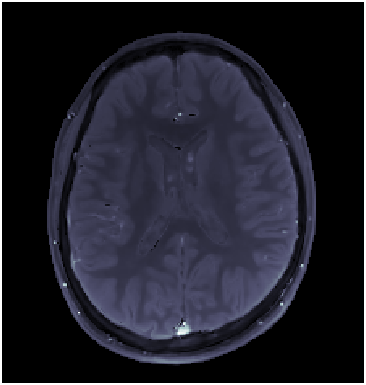}}
			%  \vspace{1.5cm}
			%. \centerline{(c) Result 4}\medskip 			% To add a manual subfigure caption
		\end{minipage}
		\\ % ----------------------------------------------------
		\begin{minipage}[b]{6.35cm}
			\centering
			\vspace{1.0mm}
			\centerline{\includegraphics[scale=0.55]{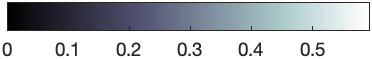}}
			\vspace{2.0mm}
			\centerline{(e) GT: Ch2}\medskip 		% To add a manual subfigure caption
		\end{minipage}
		\hfill
		\begin{minipage}[b]{-6.35cm}
			\centering
			\vspace{1.0mm}
			\centerline{\includegraphics[scale=0.55]{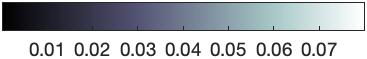}}
			\vspace{2.0mm}
			\centerline{(f) SVD-MRF: Ch2}\medskip 		% To add a manual subfigure caption
		\end{minipage}
		\hfill
		\begin{minipage}[b]{-6.35cm}
			\centering
			\vspace{1.0mm}
			\centerline{\includegraphics[scale=0.55]{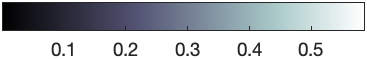}}
			\vspace{2.0mm}
			\centerline{(g) LRTV: Ch2}\medskip 		% To add a manual subfigure caption
		\end{minipage}
		\hfill
		\begin{minipage}[b]{-6.4cm}
			\centering
			\vspace{1.0mm}
			\centerline{\includegraphics[scale=0.55]{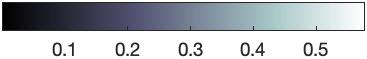}}
			\vspace{2.0mm}
			\centerline{(h) PnP-ADMM: Ch2}\medskip 		% To add a manual subfigure caption
		\end{minipage}
		%
		% ------------------------------------------------ Ch 2 --------------------------------------------------------
		%
		\\
		\vspace{3.5mm} %1.5mm, 2.5mm
		%
		% ------------------------------------------------ Ch 3 --------------------------------------------------------
		%
		\begin{minipage}[b]{6.4cm}
			\centering
			\centerline{\includegraphics[scale=0.55]{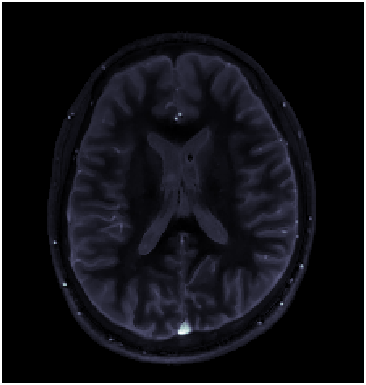}}
			%\vspace{-0.6mm}
			%  \centerline{(b) Results 3}\medskip 		% To add a manual subfigure caption
		\end{minipage}
		\hfill
		\begin{minipage}[b]{-6.4cm}
			\centering
			\centerline{\includegraphics[scale=0.55]{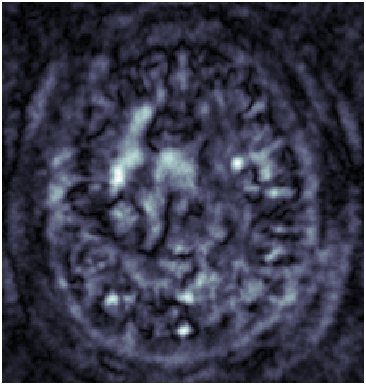}}
			%  \vspace{1.5cm}
			%  \centerline{(b) Results 3}\medskip 		% To add a manual subfigure caption
		\end{minipage}
		\hfill
		\begin{minipage}[b]{-6.4cm}
			\centering
			\centerline{\includegraphics[scale=0.55]{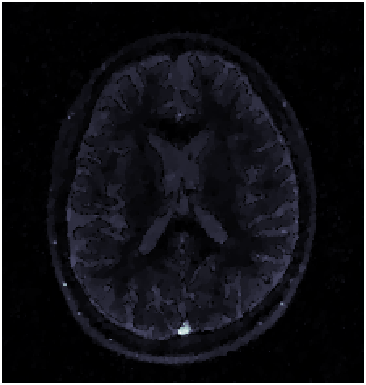}}
			%  \vspace{1.5cm}
			%. \centerline{(c) Result 4}\medskip 			% To add a manual subfigure caption
		\end{minipage}
		\hfill
		\begin{minipage}[b]{-6.4cm}
			\centering
			\centerline{\includegraphics[scale=0.55]{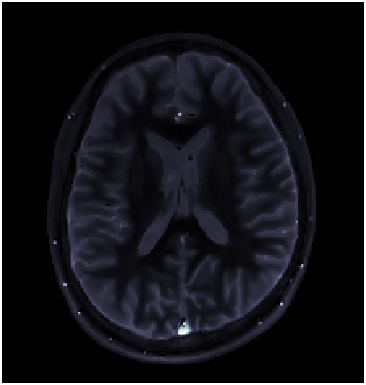}}
			%  \vspace{1.5cm}
			%. \centerline{(c) Result 4}\medskip 			% To add a manual subfigure caption
		\end{minipage}
		\\ % ----------------------------------------------------
		\begin{minipage}[b]{6.35cm}
			\centering
			\vspace{1.0mm}
			\centerline{\includegraphics[scale=0.55]{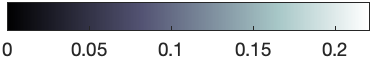}}
			\vspace{2.0mm}
			\centerline{(i) GT: Ch3}\medskip 		% To add a manual subfigure caption
		\end{minipage}
		\hfill
		\begin{minipage}[b]{-6.35cm}
			\centering
			\vspace{1.0mm}
			\centerline{\includegraphics[scale=0.55]{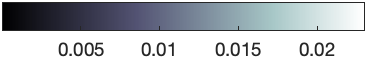}}
			\vspace{2.0mm}
			\centerline{(j) SVD-MRF: Ch3}\medskip 		% To add a manual subfigure caption
		\end{minipage}
		\hfill
		\begin{minipage}[b]{-6.35cm}
			\centering
			\vspace{1.0mm}
			\centerline{\includegraphics[scale=0.55]{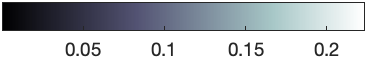}}
			\vspace{2.0mm}
			\centerline{(k) LRTV: Ch3}\medskip 		% To add a manual subfigure caption
		\end{minipage}
		\hfill
		\begin{minipage}[b]{-6.4cm}
			\centering
			\vspace{1.0mm}
			\centerline{\includegraphics[scale=0.55]{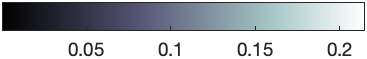}}
			\vspace{2.0mm}
			\centerline{(l) PnP-ADMM: Ch3}\medskip 		% To add a manual subfigure caption
		\end{minipage}
		%
		% ------------------------------------------------ Ch 3 --------------------------------------------------------
		%
		\\ 	% Need this here otherwise things will shift slightly
		\vspace{0.5cm}
	\caption{A visual comparison of the TSMIs obtained using the Spiral Subsampling pattern for channels 1, 2 and 3, for slice 10.}
	\label{fig:supplementary_tsmi_fig_spiral_1}
\end{center}

}]
% ---------- Supplementary Material - Spiral TSMIs Fig - 1 (Camera Ready - Title Page) ----------
%
% ---------- Supplementary Material - Spiral TSMIs Fig - 2 (Camera Ready) ----------
\twocolumn[{
\renewcommand\twocolumn[1][]{#1}
% Reference: https://tex.stackexchange.com/questions/53966/how-to-place-a-banner-image-at-the-top-of-a-paper?rq=1
% Reference: https://tex.stackexchange.com/questions/53979/span-columns-with-a-center-environment/53984#53984
% Reference: https://tex.stackexchange.com/questions/55764/input-a-figure-between-title-and-body-in-twocolumn-form
\makeatletter 	
\newcommand*\captiontype[1]{\def\@captype{#1}} 
\makeatother 

\begin{center}
	\captiontype{figure}
		\vspace{2.0cm} 	% 2.0cm, 1.7cm
		%
		% ------------------------------------------------ Ch 4 --------------------------------------------------------
		%
		\begin{minipage}[b]{6.4cm}
			\centering
			\centerline{\includegraphics[scale=0.55]{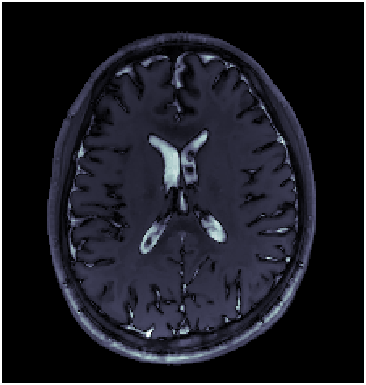}}
			%\vspace{-0.6mm}
			%  \centerline{(b) Results 3}\medskip 		% To add a manual subfigure caption
		\end{minipage}
		\hfill
		\begin{minipage}[b]{-6.4cm}
			\centering
			\centerline{\includegraphics[scale=0.55]{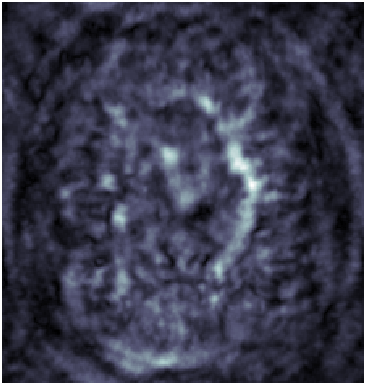}}
			%  \vspace{1.5cm}
			%  \centerline{(b) Results 3}\medskip 		% To add a manual subfigure caption
		\end{minipage}
		\hfill
		\begin{minipage}[b]{-6.4cm}
			\centering
			\centerline{\includegraphics[scale=0.55]{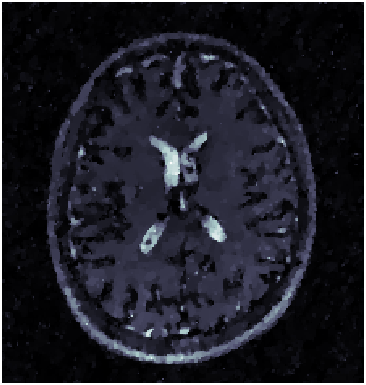}}
			%  \vspace{1.5cm}
			%. \centerline{(c) Result 4}\medskip 			% To add a manual subfigure caption
		\end{minipage}
		\hfill
		\begin{minipage}[b]{-6.4cm}
			\centering
			\centerline{\includegraphics[scale=0.55]{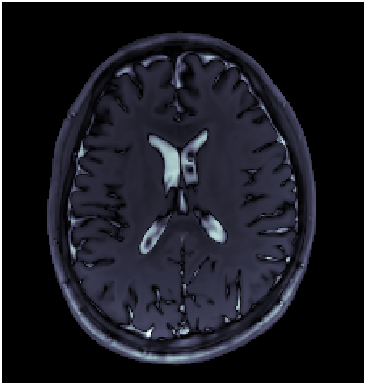}}
			%  \vspace{1.5cm}
			%. \centerline{(c) Result 4}\medskip 			% To add a manual subfigure caption
		\end{minipage}
		\\ % ----------------------------------------------------
		\begin{minipage}[b]{6.5cm}
			\centering
			\vspace{1.0mm}
			\centerline{\includegraphics[scale=0.55]{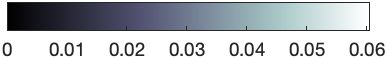}}
			\vspace{2.0mm}
			\centerline{(a) GT: Ch4}\medskip 		% To add a manual subfigure caption
		\end{minipage}
		\hfill
		\begin{minipage}[b]{-6.525cm}
			\centering
			\vspace{1.0mm}
			\centerline{\includegraphics[scale=0.55]{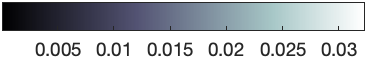}}
			\vspace{2.0mm}
			\centerline{(b) SVD-MRF: Ch4}\medskip 		% To add a manual subfigure caption
		\end{minipage}
		\hfill
		\begin{minipage}[b]{-6.45cm}
			\centering
			\vspace{1.0mm}
			\centerline{\includegraphics[scale=0.55]{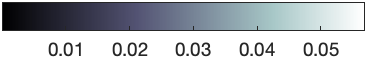}}
			\vspace{2.0mm}
			\centerline{(c) LRTV: Ch4}\medskip 		% To add a manual subfigure caption
		\end{minipage}
		\hfill
		\begin{minipage}[b]{-6.3cm}
			\centering
			\vspace{1.0mm}
			\centerline{\includegraphics[scale=0.55]{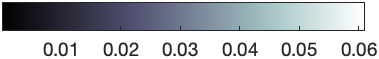}}
			\vspace{2.0mm}
			\centerline{(d) PnP-ADMM: Ch4}\medskip 		% To add a manual subfigure caption
		\end{minipage}
		%
		% ------------------------------------------------ Ch 4 --------------------------------------------------------
		%
		\\
		\vspace{3.5mm} %1.5, 2.5mm
		%
		% ------------------------------------------------ Ch 5 --------------------------------------------------------
		%
		\begin{minipage}[b]{6.4cm}
			\centering
			\centerline{\includegraphics[scale=0.55]{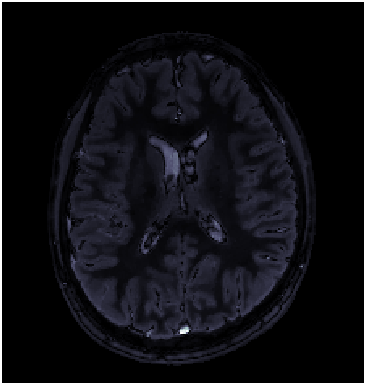}}
			%\vspace{-0.6mm}
			%  \centerline{(b) Results 3}\medskip 		% To add a manual subfigure caption
		\end{minipage}
		\hfill
		\begin{minipage}[b]{-6.4cm}
			\centering
			\centerline{\includegraphics[scale=0.55]{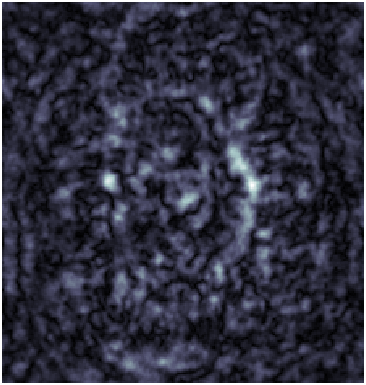}}
			%  \vspace{1.5cm}
			%  \centerline{(b) Results 3}\medskip 		% To add a manual subfigure caption
		\end{minipage}
		\hfill
		\begin{minipage}[b]{-6.4cm}
			\centering
			\centerline{\includegraphics[scale=0.55]{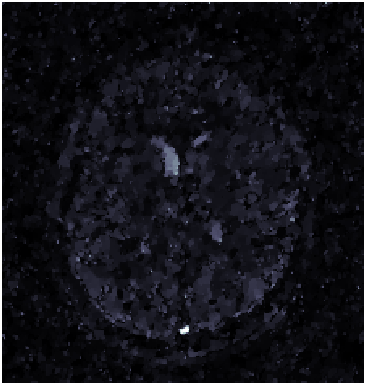}}
			%  \vspace{1.5cm}
			%. \centerline{(c) Result 4}\medskip 			% To add a manual subfigure caption
		\end{minipage}
		\hfill
		\begin{minipage}[b]{-6.4cm}
			\centering
			\centerline{\includegraphics[scale=0.55]{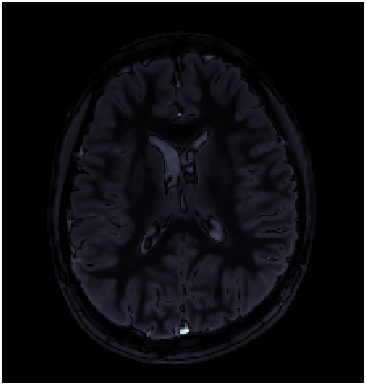}}
			%  \vspace{1.5cm}
			%. \centerline{(c) Result 4}\medskip 			% To add a manual subfigure caption
		\end{minipage}
		\\ % ----------------------------------------------------
		\begin{minipage}[b]{6.35cm}
			\centering
			\vspace{1.0mm}
			\centerline{\includegraphics[scale=0.55]{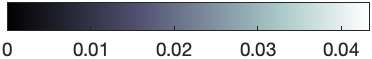}}
			\vspace{2.0mm}
			\centerline{(e) GT: Ch5}\medskip 		% To add a manual subfigure caption
		\end{minipage}
		\hfill
		\begin{minipage}[b]{-6.325cm}
			\centering
			\vspace{1.0mm}
			\centerline{\includegraphics[scale=0.55]{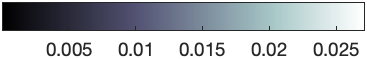}}
			\vspace{2.0mm}
			\centerline{(f) SVD-MRF: Ch5}\medskip 		% To add a manual subfigure caption
		\end{minipage}
		\hfill
		\begin{minipage}[b]{-6.35cm}
			\centering
			\vspace{1.0mm}
			\centerline{\includegraphics[scale=0.55]{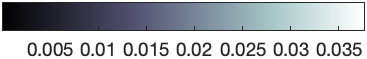}}
			\vspace{2.0mm}
			\centerline{(g) LRTV: Ch5}\medskip 		% To add a manual subfigure caption
		\end{minipage}
		\hfill
		\begin{minipage}[b]{-6.4cm}
			\centering
			\vspace{1.0mm}
			\centerline{\includegraphics[scale=0.55]{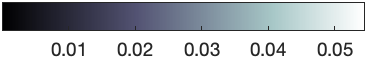}}
			\vspace{2.0mm}
			\centerline{(h) PnP-ADMM: Ch5}\medskip 		% To add a manual subfigure caption
		\end{minipage}
		%
		% ------------------------------------------------ Ch 5 --------------------------------------------------------
		%
		\\
		\vspace{3.5mm} %1.5mm, 2.5mm
		%
		% ------------------------------------------------ Ch 6 --------------------------------------------------------
		%
		\begin{minipage}[b]{6.4cm}
			\centering
			\centerline{\includegraphics[scale=0.55]{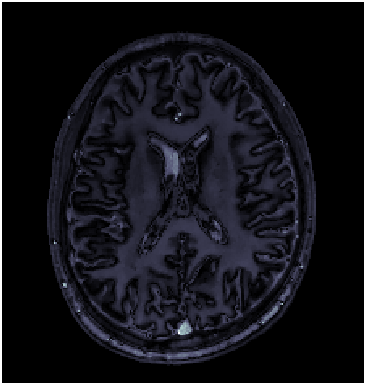}}
			%\vspace{-0.6mm}
			%  \centerline{(b) Results 3}\medskip 		% To add a manual subfigure caption
		\end{minipage}
		\hfill
		\begin{minipage}[b]{-6.4cm}
			\centering
			\centerline{\includegraphics[scale=0.55]{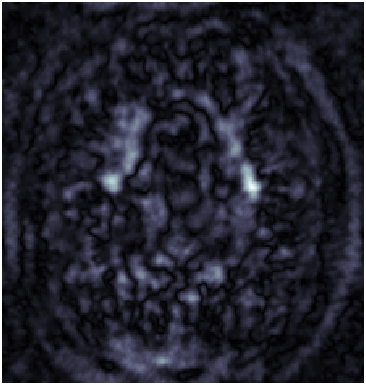}}
			%  \vspace{1.5cm}
			%  \centerline{(b) Results 3}\medskip 		% To add a manual subfigure caption
		\end{minipage}
		\hfill
		\begin{minipage}[b]{-6.4cm}
			\centering
			\centerline{\includegraphics[scale=0.55]{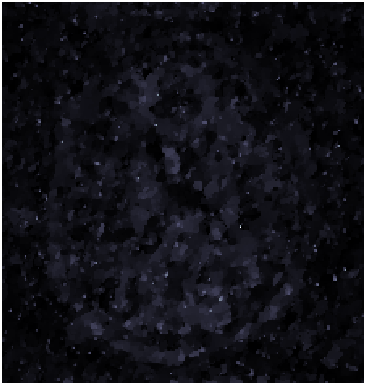}}
			%  \vspace{1.5cm}
			%. \centerline{(c) Result 4}\medskip 			% To add a manual subfigure caption
		\end{minipage}
		\hfill
		\begin{minipage}[b]{-6.4cm}
			\centering
			\centerline{\includegraphics[scale=0.55]{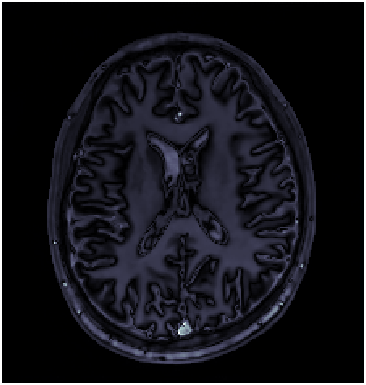}}
			%  \vspace{1.5cm}
			%. \centerline{(c) Result 4}\medskip 			% To add a manual subfigure caption
		\end{minipage}
		\\ % ----------------------------------------------------
		\begin{minipage}[b]{6.35cm}
			\centering
			\vspace{1.0mm}
			\centerline{\includegraphics[scale=0.55]{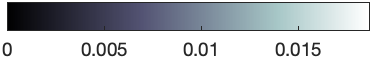}}
			\vspace{4.8mm}
			\centerline{(i) GT: Ch6}\medskip 		% To add a manual subfigure caption
		\end{minipage}
		\hfill
		\begin{minipage}[b]{-6.325cm}
			\centering
			\vspace{1.0mm}
			\centerline{\includegraphics[scale=0.55]{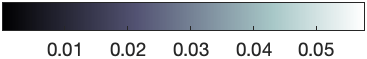}}
			\vspace{4.7mm}
			\centerline{(j) SVD-MRF: Ch6}\medskip 		% To add a manual subfigure caption
		\end{minipage}
		\hfill
		\begin{minipage}[b]{-6.35cm}
			\centering
			\vspace{1.0mm}
			\centerline{\includegraphics[scale=0.55]{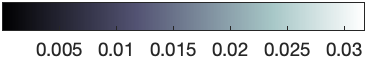}}
			\vspace{4.8mm}
			\centerline{(k) LRTV: Ch6}\medskip 		% To add a manual subfigure caption
		\end{minipage}
		\hfill
		\begin{minipage}[b]{-6.4cm}
			\centering
			\vspace{1.0mm}
			\centerline{\includegraphics[scale=0.55]{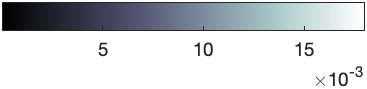}}
			\vspace{2.0mm}
			\centerline{(l) PnP-ADMM: Ch6}\medskip 		% To add a manual subfigure caption
		\end{minipage}
		%
		% ------------------------------------------------ Ch 6 --------------------------------------------------------
		%
		\\ 	% Need this here otherwise things will shift slightly
		\vspace{0.5cm}
	\caption{A visual comparison of the TSMIs obtained using the Spiral Subsampling pattern for channels 4, 5 and 6, for slice 10.}
	\label{fig:supplementary_tsmi_fig_spiral_2}
\end{center}

}]
% ---------- Supplementary Material - Spiral TSMIs Fig - 2 (Camera Ready) ----------
% ---------- Supplementary Material - Spiral TSMIs Fig - 3 (Camera Ready) ----------
\twocolumn[{
\renewcommand\twocolumn[1][]{#1}
% Reference: https://tex.stackexchange.com/questions/53966/how-to-place-a-banner-image-at-the-top-of-a-paper?rq=1
% Reference: https://tex.stackexchange.com/questions/53979/span-columns-with-a-center-environment/53984#53984
% Reference: https://tex.stackexchange.com/questions/55764/input-a-figure-between-title-and-body-in-twocolumn-form
\makeatletter 	
\newcommand*\captiontype[1]{\def\@captype{#1}} 
\makeatother 

\begin{center}
	\captiontype{figure}
		\vspace{1.7cm} 	% 1.7cm, 1.75cm
		%
		% ------------------------------------------------ Ch 7 --------------------------------------------------------
		%
		\begin{minipage}[b]{6.4cm}
			\centering
			\centerline{\includegraphics[scale=0.55]{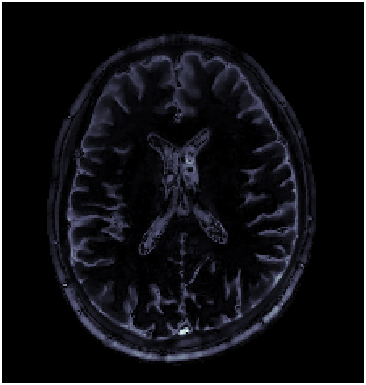}}
			%\vspace{-0.6mm}
			%  \centerline{(b) Results 3}\medskip 		% To add a manual subfigure caption
		\end{minipage}
		\hfill
		\begin{minipage}[b]{-6.4cm}
			\centering
			\centerline{\includegraphics[scale=0.55]{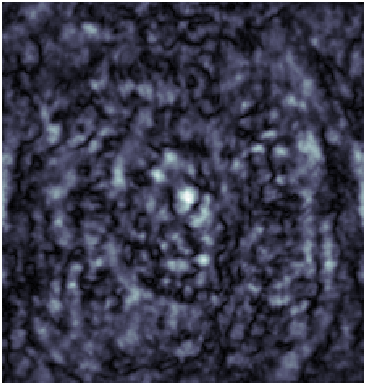}}
			%  \vspace{1.5cm}
			%  \centerline{(b) Results 3}\medskip 		% To add a manual subfigure caption
		\end{minipage}
		\hfill
		\begin{minipage}[b]{-6.4cm}
			\centering
			\centerline{\includegraphics[scale=0.55]{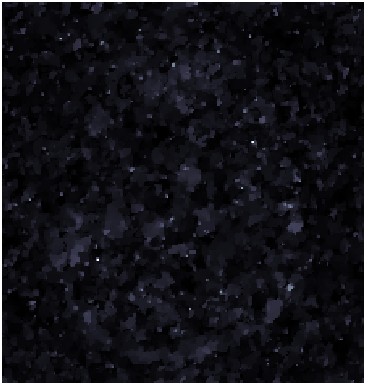}}
			%  \vspace{1.5cm}
			%. \centerline{(c) Result 4}\medskip 			% To add a manual subfigure caption
		\end{minipage}
		\hfill
		\begin{minipage}[b]{-6.4cm}
			\centering
			\centerline{\includegraphics[scale=0.55]{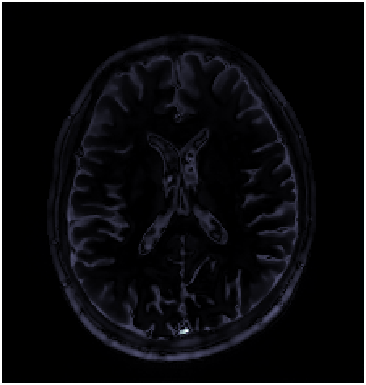}}
			%  \vspace{1.5cm}
			%. \centerline{(c) Result 4}\medskip 			% To add a manual subfigure caption
		\end{minipage}
		\\ % ----------------------------------------------------
		\begin{minipage}[b]{6.35cm}
			\centering
			\vspace{1.0mm}
			\centerline{\includegraphics[scale=0.55]{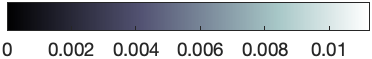}}
			\vspace{4.9mm}
			\centerline{(a) GT: Ch7}\medskip 		% To add a manual subfigure caption
		\end{minipage}
		\hfill
		\begin{minipage}[b]{-6.325cm}
			\centering
			\vspace{1.0mm}
			\centerline{\includegraphics[scale=0.55]{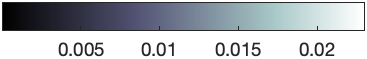}}
			\vspace{4.9mm}
			\centerline{(b) SVD-MRF: Ch7}\medskip 		% To add a manual subfigure caption
		\end{minipage}
		\hfill
		\begin{minipage}[b]{-6.375cm}
			\centering
			\vspace{1.0mm}
			\centerline{\includegraphics[scale=0.55]{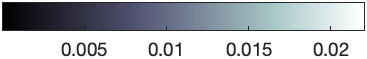}}
			\vspace{4.9mm}
			\centerline{(c) LRTV: Ch7}\medskip 		% To add a manual subfigure caption
		\end{minipage}
		\hfill
		\begin{minipage}[b]{-6.4cm}
			\centering
			\vspace{1.0mm}
			\centerline{\includegraphics[scale=0.55]{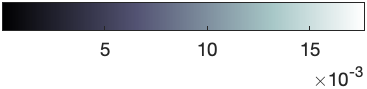}}
			\vspace{2.0mm}
			\centerline{(d) PnP-ADMM: Ch7}\medskip 		% To add a manual subfigure caption
		\end{minipage}
		%
		% ------------------------------------------------ Ch 7 --------------------------------------------------------
		%
		\\
		\vspace{3.5mm} %1.5, 2.5mm
		%
		% ------------------------------------------------ Ch 8 --------------------------------------------------------
		%
		\begin{minipage}[b]{6.4cm}
			\centering
			\centerline{\includegraphics[scale=0.55]{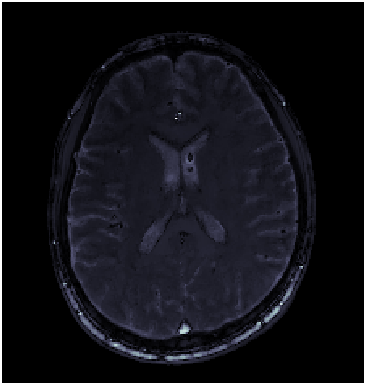}}
			%\vspace{-0.6mm}
			%  \centerline{(b) Results 3}\medskip 		% To add a manual subfigure caption
		\end{minipage}
		\hfill
		\begin{minipage}[b]{-6.4cm}
			\centering
			\centerline{\includegraphics[scale=0.55]{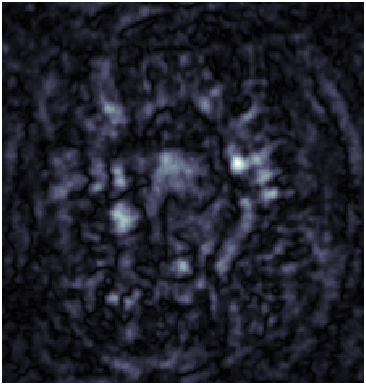}}
			%  \vspace{1.5cm}
			%  \centerline{(b) Results 3}\medskip 		% To add a manual subfigure caption
		\end{minipage}
		\hfill
		\begin{minipage}[b]{-6.4cm}
			\centering
			\centerline{\includegraphics[scale=0.55]{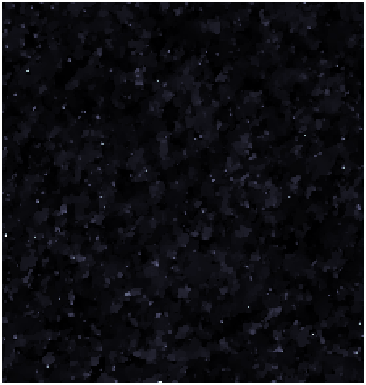}}
			%  \vspace{1.5cm}
			%. \centerline{(c) Result 4}\medskip 			% To add a manual subfigure caption
		\end{minipage}
		\hfill
		\begin{minipage}[b]{-6.4cm}
			\centering
			\centerline{\includegraphics[scale=0.55]{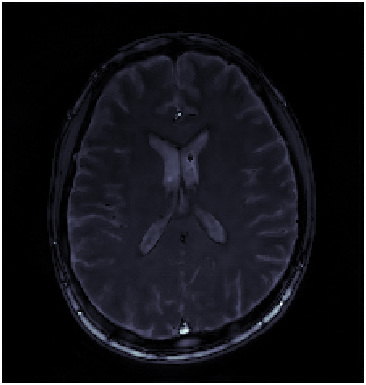}}
			%  \vspace{1.5cm}
			%. \centerline{(c) Result 4}\medskip 			% To add a manual subfigure caption
		\end{minipage}
		\\ % ----------------------------------------------------
		\begin{minipage}[b]{6.35cm}
			\centering
			\vspace{1.0mm}
			\centerline{\includegraphics[scale=0.55]{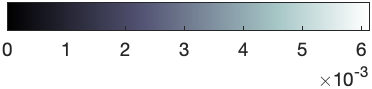}}
			\vspace{2.0mm}
			\centerline{(e) GT: Ch8}\medskip 		% To add a manual subfigure caption
		\end{minipage}
		\hfill
		\begin{minipage}[b]{-6.325cm}
			\centering
			\vspace{1.0mm}
			\centerline{\includegraphics[scale=0.55]{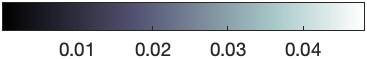}}
			\vspace{4.8mm}
			\centerline{(f) SVD-MRF: Ch8}\medskip 		% To add a manual subfigure caption
		\end{minipage}
		\hfill
		\begin{minipage}[b]{-6.175cm}
			\centering
			\vspace{1.0mm}
			\centerline{\includegraphics[scale=0.55]{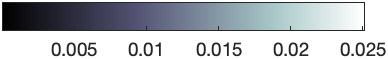}}
			\vspace{4.8mm}
			\centerline{(g) LRTV: Ch8}\medskip 		% To add a manual subfigure caption
		\end{minipage}
		\hfill
		\begin{minipage}[b]{-6.4cm}
			\centering
			\vspace{1.0mm}
			\centerline{\includegraphics[scale=0.55]{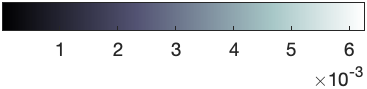}}
			\vspace{2.0mm}
			\centerline{(h) PnP-ADMM: Ch8}\medskip 		% To add a manual subfigure caption
		\end{minipage}
		%
		% ------------------------------------------------ Ch 8 --------------------------------------------------------
		%
		\\
		\vspace{3.5mm} %1.5, 2.5mm
		%
		% ------------------------------------------------ Ch 9 --------------------------------------------------------
		%
		\begin{minipage}[b]{6.4cm}
			\centering
			\centerline{\includegraphics[scale=0.55]{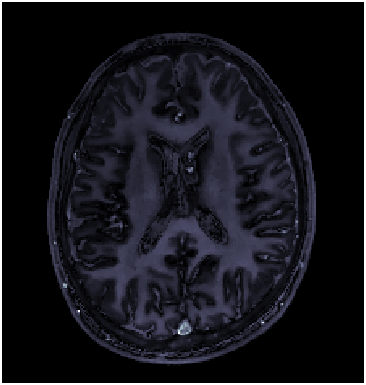}}
			%\vspace{-0.6mm}
			%  \centerline{(b) Results 3}\medskip 		% To add a manual subfigure caption
		\end{minipage}
		\hfill
		\begin{minipage}[b]{-6.4cm}
			\centering
			\centerline{\includegraphics[scale=0.55]{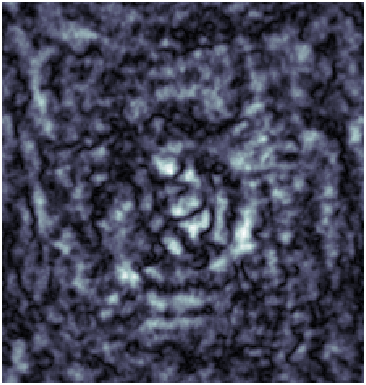}}
			%  \vspace{1.5cm}
			%  \centerline{(b) Results 3}\medskip 		% To add a manual subfigure caption
		\end{minipage}
		\hfill
		\begin{minipage}[b]{-6.4cm}
			\centering
			\centerline{\includegraphics[scale=0.55]{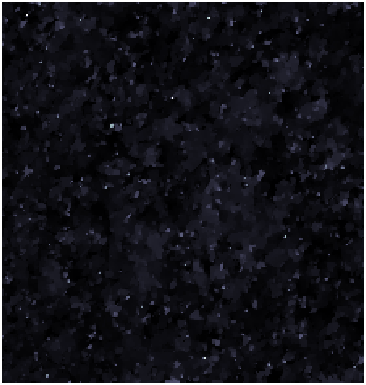}}
			%  \vspace{1.5cm}
			%. \centerline{(c) Result 4}\medskip 			% To add a manual subfigure caption
		\end{minipage}
		\hfill
		\begin{minipage}[b]{-6.4cm}
			\centering
			\centerline{\includegraphics[scale=0.55]{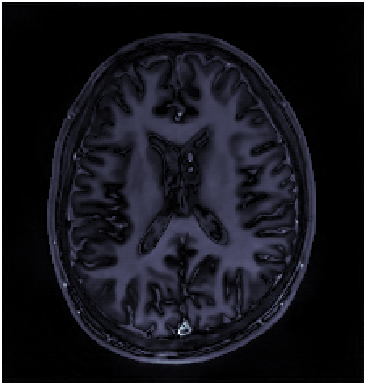}}
			%  \vspace{1.5cm}
			%. \centerline{(c) Result 4}\medskip 			% To add a manual subfigure caption
		\end{minipage}
		\\ % ----------------------------------------------------
		\begin{minipage}[b]{6.35cm}
			\centering
			\vspace{1.0mm}
			\centerline{\includegraphics[scale=0.55]{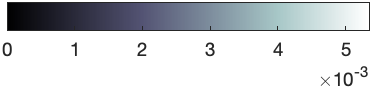}}
			\vspace{2.0mm}
			\centerline{(i) GT: Ch9}\medskip 		% To add a manual subfigure caption
		\end{minipage}
		\hfill
		\begin{minipage}[b]{-6.325cm}
			\centering
			\vspace{1.0mm}
			\centerline{\includegraphics[scale=0.55]{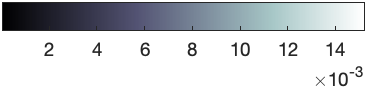}}
			\vspace{2.0mm}
			\centerline{(j) SVD-MRF: Ch9}\medskip 		% To add a manual subfigure caption
		\end{minipage}
		\hfill
		\begin{minipage}[b]{-6.35cm}
			\centering
			\vspace{1.0mm}
			\centerline{\includegraphics[scale=0.55]{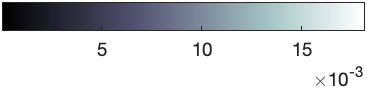}}
			\vspace{2.0mm}
			\centerline{(k) LRTV: Ch9}\medskip 		% To add a manual subfigure caption
		\end{minipage}
		\hfill
		\begin{minipage}[b]{-6.375cm}
			\centering
			\vspace{1.0mm}
			\centerline{\includegraphics[scale=0.55]{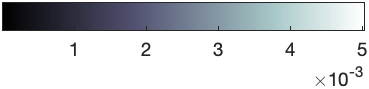}}
			\vspace{2.0mm}
			\centerline{(l) PnP-ADMM: Ch9}\medskip 		% To add a manual subfigure caption
		\end{minipage}
		%
		% ------------------------------------------------ Ch 9 --------------------------------------------------------
		%
		\\ 	% Need this here otherwise things will shift slightly
		\vspace{0.5cm}
	\caption{A visual comparison of the TSMIs obtained using the Spiral Subsampling pattern for channels 7, 8 and 9, for slice 10.}
	\label{fig:supplementary_tsmi_fig_spiral_3}
\end{center}

}]
% ---------- Supplementary Material - Spiral TSMIs Fig - 3 (Camera Ready) ----------
% ---------- Supplementary Material - Spiral TSMIs Fig - 4 (Camera Ready) ----------
\twocolumn[{
\renewcommand\twocolumn[1][]{#1}
% Reference: https://tex.stackexchange.com/questions/53966/how-to-place-a-banner-image-at-the-top-of-a-paper?rq=1
% Reference: https://tex.stackexchange.com/questions/53979/span-columns-with-a-center-environment/53984#53984
% Reference: https://tex.stackexchange.com/questions/55764/input-a-figure-between-title-and-body-in-twocolumn-form
\makeatletter 
\newcommand*\captiontype[1]{\def\@captype{#1}} 
\makeatother 

\begin{center}
	\captiontype{figure}
		\vspace{6.5cm} 	% 6.5cm,
		%
		% ------------------------------------------------ Ch 10 --------------------------------------------------------
		%
		\begin{minipage}[b]{6.4cm}
			\centering
			\centerline{\includegraphics[scale=0.55]{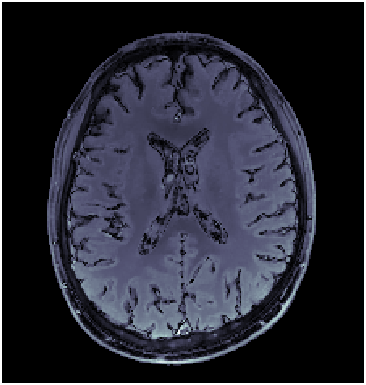}}
			%\vspace{-0.6mm}
			%  \centerline{(b) Results 3}\medskip 		% To add a manual subfigure caption
		\end{minipage}
		\hfill
		\begin{minipage}[b]{-6.4cm}
			\centering
			\centerline{\includegraphics[scale=0.55]{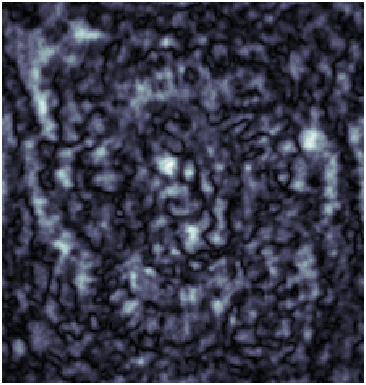}}
			%  \vspace{1.5cm}
			%  \centerline{(b) Results 3}\medskip 		% To add a manual subfigure caption
		\end{minipage}
		\hfill
		\begin{minipage}[b]{-6.4cm}
			\centering
			\centerline{\includegraphics[scale=0.55]{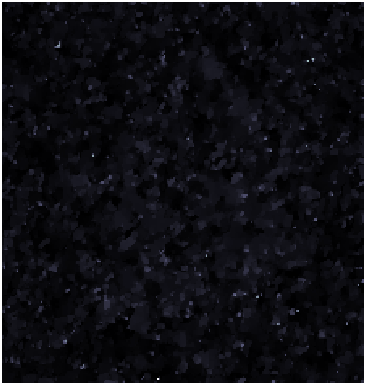}}
			%  \vspace{1.5cm}
			%. \centerline{(c) Result 4}\medskip 			% To add a manual subfigure caption
		\end{minipage}
		\hfill
		\begin{minipage}[b]{-6.4cm}
			\centering
			\centerline{\includegraphics[scale=0.55]{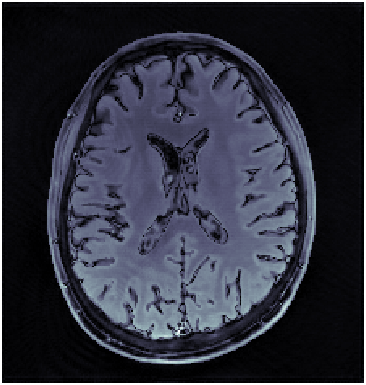}}
			%  \vspace{1.5cm}
			%. \centerline{(c) Result 4}\medskip 			% To add a manual subfigure caption
		\end{minipage}
		\\ % ----------------------------------------------------
		\begin{minipage}[b]{6.35cm}
			\centering
			\vspace{1.0mm}
			\centerline{\includegraphics[scale=0.55]{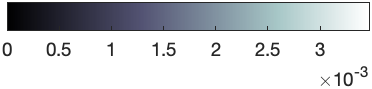}}
			\vspace{2.0mm}
			\centerline{(a) GT: Ch10}\medskip 		% To add a manual subfigure caption
		\end{minipage}
		\hfill
		\begin{minipage}[b]{-6.325cm}
			\centering
			\vspace{1.0mm}
			\centerline{\includegraphics[scale=0.55]{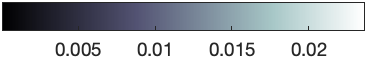}}
			\vspace{4.65mm}
			\centerline{(b) SVD-MRF: Ch10}\medskip 		% To add a manual subfigure caption
		\end{minipage}
		\hfill
		\begin{minipage}[b]{-6.375cm}
			\centering
			\vspace{1.0mm}
			\centerline{\includegraphics[scale=0.55]{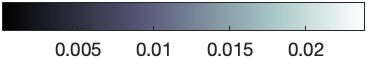}}
			\vspace{4.8mm}
			\centerline{(c) LRTV: Ch10}\medskip 		% To add a manual subfigure caption
		\end{minipage}
		\hfill
		\begin{minipage}[b]{-6.4cm}
			\centering
			\vspace{1.0mm}
			\centerline{\includegraphics[scale=0.55]{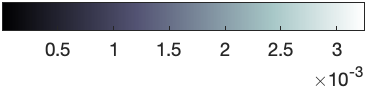}}
			\vspace{2.0mm}
			\centerline{(d) PnP-ADMM: Ch10}\medskip 		% To add a manual subfigure caption
		\end{minipage}
		%
		% ------------------------------------------------ Ch 10 --------------------------------------------------------
		%
		\\ 	% Need this here otherwise things will shift slightly
		\vspace{0.5cm}
	\caption{A visual comparison of the TSMIs obtained using the Spiral Subsampling pattern for channel 10, for slice 10.}
	\label{fig:supplementary_tsmi_fig_spiral_4}
\end{center}

}]
% ---------- Supplementary Material - Spiral TSMIs Fig - 4 (Camera Ready) ----------
%
% ---------------------------------------------------------- Spiral ----------------------------------------------------------
%
% ---------------------------------------------------------- EPI ----------------------------------------------------------
%
% ---------- Supplementary Material - EPI TSMIs Fig - 1 (Camera Ready) ----------
\twocolumn[{
\renewcommand\twocolumn[1][]{#1}
% Reference: https://tex.stackexchange.com/questions/53966/how-to-place-a-banner-image-at-the-top-of-a-paper?rq=1
% Reference: https://tex.stackexchange.com/questions/53979/span-columns-with-a-center-environment/53984#53984
% Reference: https://tex.stackexchange.com/questions/55764/input-a-figure-between-title-and-body-in-twocolumn-form
\makeatletter 
\newcommand*\captiontype[1]{\def\@captype{#1}} 
\makeatother 

\begin{center}
	\captiontype{figure}
		\vspace{1.7cm} 	% 1.7cm, 1.75cm
		%
		% ------------------------------------------------ Ch 1 --------------------------------------------------------
		%
		\begin{minipage}[b]{6.4cm}
			\centering
			\centerline{\includegraphics[scale=0.55]{./figures/tsmis/ground_truth/tsmi_ground_truth_ch1.png}}
			%\vspace{-0.6mm}
			%  \centerline{(b) Results 3}\medskip 		% To add a manual subfigure caption
		\end{minipage}
		\hfill
		\begin{minipage}[b]{-6.4cm}
			\centering
			\centerline{\includegraphics[scale=0.55]{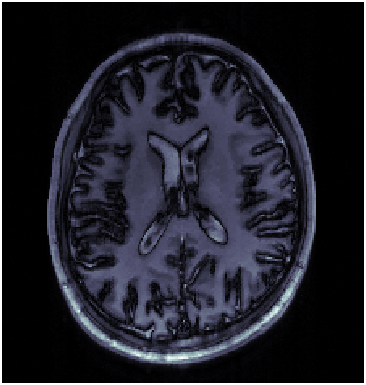}}
			%  \vspace{1.5cm}
			%  \centerline{(b) Results 3}\medskip 		% To add a manual subfigure caption
		\end{minipage}
		\hfill
		\begin{minipage}[b]{-6.4cm}
			\centering
			\centerline{\includegraphics[scale=0.55]{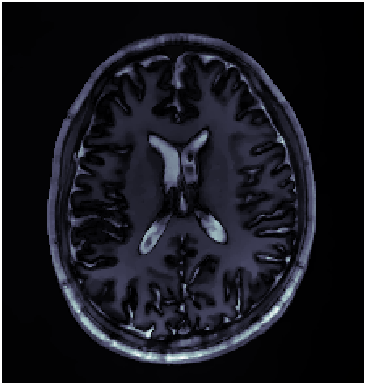}}
			%  \vspace{1.5cm}
			%. \centerline{(c) Result 4}\medskip 			% To add a manual subfigure caption
		\end{minipage}
		\hfill
		\begin{minipage}[b]{-6.4cm}
			\centering
			\centerline{\includegraphics[scale=0.55]{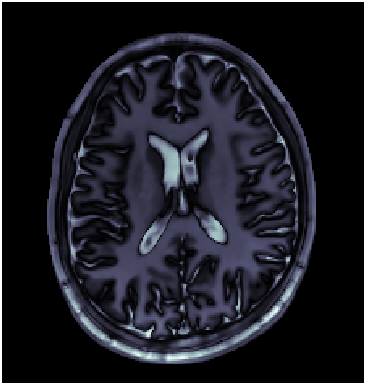}}
			%  \vspace{1.5cm}
			%. \centerline{(c) Result 4}\medskip 			% To add a manual subfigure caption
		\end{minipage}
		\\ % ----------------------------------------------------
		\begin{minipage}[b]{6.4cm}
			\centering
			\vspace{1.0mm}
			\centerline{\includegraphics[scale=0.55]{./figures/tsmis/ground_truth/tsmi_ground_truth_ch1_colorbar_horizontal_south.png}}
			\vspace{4.8mm}
			\centerline{(a) GT: Ch1}\medskip 		% To add a manual subfigure caption
		\end{minipage}
		\hfill
		\begin{minipage}[b]{-6.4cm}
			\centering
			\vspace{1.0mm}
			\centerline{\includegraphics[scale=0.55]{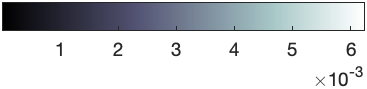}}
			\vspace{2.0mm}
			\centerline{(b) SVD-MRF: Ch1}\medskip 		% To add a manual subfigure caption
		\end{minipage}
		\hfill
		\begin{minipage}[b]{-6.4cm}
			\centering
			\vspace{1.0mm}
			\centerline{\includegraphics[scale=0.55]{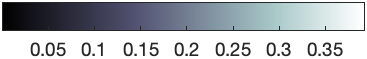}}
			\vspace{4.8mm}
			\centerline{(c) LRTV: Ch1}\medskip 		% To add a manual subfigure caption
		\end{minipage}
		\hfill
		\begin{minipage}[b]{-6.3cm}
			\centering
			\vspace{1.0mm}
			\centerline{\includegraphics[scale=0.55]{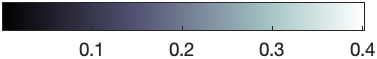}}
			\vspace{4.8mm}
			\centerline{(d) PnP-ADMM: Ch1}\medskip 		% To add a manual subfigure caption
		\end{minipage}
		%
		% ------------------------------------------------ Ch 1 --------------------------------------------------------
		%
		\\
		\vspace{3.5mm} %1.5, 2.5mm
		%
		% ------------------------------------------------ Ch 2 --------------------------------------------------------
		%
		\begin{minipage}[b]{6.4cm}
			\centering
			\centerline{\includegraphics[scale=0.55]{./figures/tsmis/ground_truth/tsmi_ground_truth_ch2.png}}
			%\vspace{-0.6mm}
			%  \centerline{(b) Results 3}\medskip 		% To add a manual subfigure caption
		\end{minipage}
		\hfill
		\begin{minipage}[b]{-6.4cm}
			\centering
			\centerline{\includegraphics[scale=0.55]{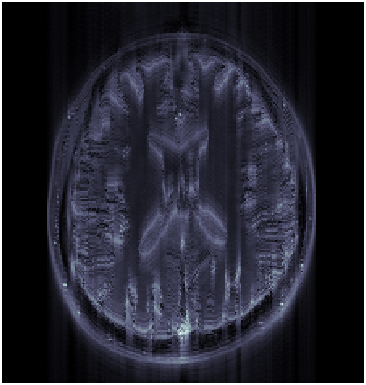}}
			%  \vspace{1.5cm}
			%  \centerline{(b) Results 3}\medskip 		% To add a manual subfigure caption
		\end{minipage}
		\hfill
		\begin{minipage}[b]{-6.4cm}
			\centering
			\centerline{\includegraphics[scale=0.55]{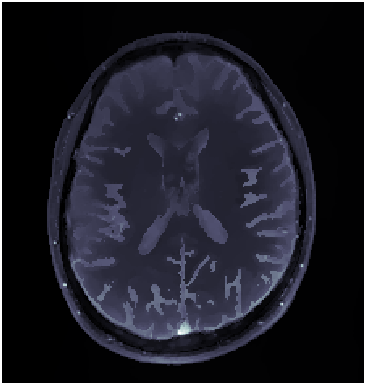}}
			%  \vspace{1.5cm}
			%. \centerline{(c) Result 4}\medskip 			% To add a manual subfigure caption
		\end{minipage}
		\hfill
		\begin{minipage}[b]{-6.4cm}
			\centering
			\centerline{\includegraphics[scale=0.55]{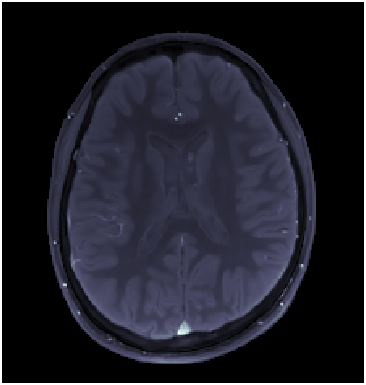}}
			%  \vspace{1.5cm}
			%. \centerline{(c) Result 4}\medskip 			% To add a manual subfigure caption
		\end{minipage}
		\\ % ----------------------------------------------------
		\begin{minipage}[b]{6.36cm}
			\centering
			\vspace{1.0mm}
			\centerline{\includegraphics[scale=0.55]{./figures/tsmis/ground_truth/tsmi_ground_truth_ch2_colorbar_horizontal_south.png}}
			\vspace{4.8mm}
			\centerline{(e) GT: Ch2}\medskip 		% To add a manual subfigure caption
		\end{minipage}
		\hfill
		\begin{minipage}[b]{-6.34cm}
			\centering
			\vspace{1.0mm}
			\centerline{\includegraphics[scale=0.55]{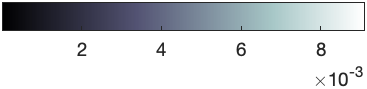}}
			\vspace{2.0mm}
			\centerline{(f) SVD-MRF: Ch2}\medskip 		% To add a manual subfigure caption
		\end{minipage}
		\hfill
		\begin{minipage}[b]{-6.375cm}
			\centering
			\vspace{1.0mm}
			\centerline{\includegraphics[scale=0.55]{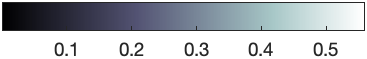}}
			\vspace{4.7mm}
			\centerline{(g) LRTV: Ch2}\medskip 		% To add a manual subfigure caption
		\end{minipage}
		\hfill
		\begin{minipage}[b]{-6.29cm}
			\centering
			\vspace{1.0mm}
			\centerline{\includegraphics[scale=0.55]{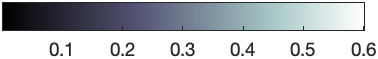}}
			\vspace{4.8mm}
			\centerline{(h) PnP-ADMM: Ch2}\medskip 		% To add a manual subfigure caption
		\end{minipage}
		%
		% ------------------------------------------------ Ch 2 --------------------------------------------------------
		%
		\\
		\vspace{3.5mm} %1.5mm, 2.5mm
		%
		% ------------------------------------------------ Ch 3 --------------------------------------------------------
		%
		\begin{minipage}[b]{6.4cm}
			\centering
			\centerline{\includegraphics[scale=0.55]{./figures/tsmis/ground_truth/tsmi_ground_truth_ch3.png}}
			%\vspace{-0.6mm}
			%  \centerline{(b) Results 3}\medskip 		% To add a manual subfigure caption
		\end{minipage}
		\hfill
		\begin{minipage}[b]{-6.4cm}
			\centering
			\centerline{\includegraphics[scale=0.55]{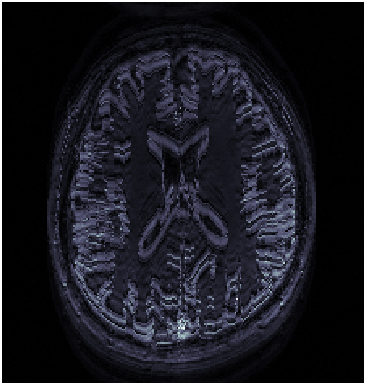}}
			%  \vspace{1.5cm}
			%  \centerline{(b) Results 3}\medskip 		% To add a manual subfigure caption
		\end{minipage}
		\hfill
		\begin{minipage}[b]{-6.4cm}
			\centering
			\centerline{\includegraphics[scale=0.55]{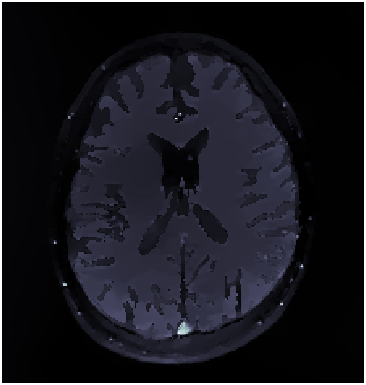}}
			%  \vspace{1.5cm}
			%. \centerline{(c) Result 4}\medskip 			% To add a manual subfigure caption
		\end{minipage}
		\hfill
		\begin{minipage}[b]{-6.4cm}
			\centering
			\centerline{\includegraphics[scale=0.55]{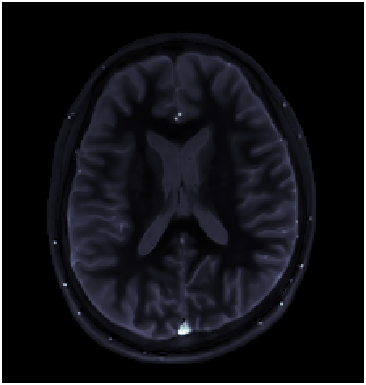}}
			%  \vspace{1.5cm}
			%. \centerline{(c) Result 4}\medskip 			% To add a manual subfigure caption
		\end{minipage}
		\\ % ----------------------------------------------------
		\begin{minipage}[b]{6.365cm}
			\centering
			\vspace{1.0mm}
			\centerline{\includegraphics[scale=0.55]{./figures/tsmis/ground_truth/tsmi_ground_truth_ch3_colorbar_horizontal_south.png}}
			\vspace{4.9mm}
			\centerline{(i) GT: Ch3}\medskip 		% To add a manual subfigure caption
		\end{minipage}
		\hfill
		\begin{minipage}[b]{-6.35cm}
			\centering
			\vspace{1.0mm}
			\centerline{\includegraphics[scale=0.55]{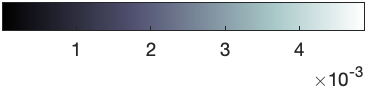}}
			\vspace{2.0mm}
			\centerline{(j) SVD-MRF: Ch3}\medskip 		% To add a manual subfigure caption
		\end{minipage}
		\hfill
		\begin{minipage}[b]{-6.375cm}
			\centering
			\vspace{1.0mm}
			\centerline{\includegraphics[scale=0.55]{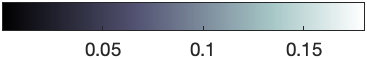}}
			\vspace{4.9mm}
			\centerline{(k) LRTV: Ch3}\medskip 		% To add a manual subfigure caption
		\end{minipage}
		\hfill
		\begin{minipage}[b]{-6.4cm}
			\centering
			\vspace{1.0mm}
			\centerline{\includegraphics[scale=0.55]{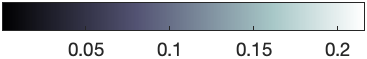}}
			\vspace{4.9mm}
			\centerline{(l) PnP-ADMM: Ch3}\medskip 		% To add a manual subfigure caption
		\end{minipage}
		%
		% ------------------------------------------------ Ch 3 --------------------------------------------------------
		%
		\\ 	% Need this here otherwise things will shift slightly
		\vspace{0.5cm}
	\caption{A visual comparison of the TSMIs obtained using the EPI Subsampling pattern for channels 1, 2 and 3, for slice 10.}
	\label{fig:supplementary_tsmi_fig_epi_1}
\end{center}

}]
% ---------- Supplementary Material - EPI TSMIs Fig - 1 (Camera Ready) ----------
% ---------- Supplementary Material - EPI TSMIs Fig - 2 (Camera Ready) ----------
\twocolumn[{
\renewcommand\twocolumn[1][]{#1}
% Reference: https://tex.stackexchange.com/questions/53966/how-to-place-a-banner-image-at-the-top-of-a-paper?rq=1
% Reference: https://tex.stackexchange.com/questions/53979/span-columns-with-a-center-environment/53984#53984
% Reference: https://tex.stackexchange.com/questions/55764/input-a-figure-between-title-and-body-in-twocolumn-form
\makeatletter 
\newcommand*\captiontype[1]{\def\@captype{#1}} 
\makeatother 

\begin{center}
	\captiontype{figure}
		\vspace{1.7cm} 	% 1.7cm, 1.75cm
		%
		% ------------------------------------------------ Ch 4 --------------------------------------------------------
		%
		\begin{minipage}[b]{6.4cm}
			\centering
			\centerline{\includegraphics[scale=0.55]{./figures/tsmis/ground_truth/tsmi_ground_truth_ch4.png}}
			%\vspace{-0.6mm}
			%  \centerline{(b) Results 3}\medskip 		% To add a manual subfigure caption
		\end{minipage}
		\hfill
		\begin{minipage}[b]{-6.4cm}
			\centering
			\centerline{\includegraphics[scale=0.55]{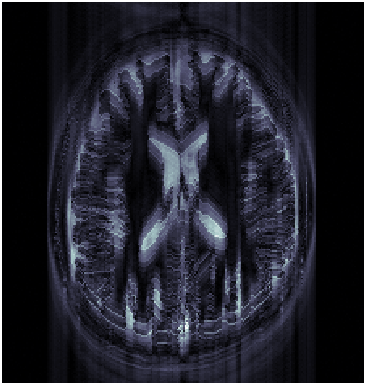}}
			%  \vspace{1.5cm}
			%  \centerline{(b) Results 3}\medskip 		% To add a manual subfigure caption
		\end{minipage}
		\hfill
		\begin{minipage}[b]{-6.4cm}
			\centering
			\centerline{\includegraphics[scale=0.55]{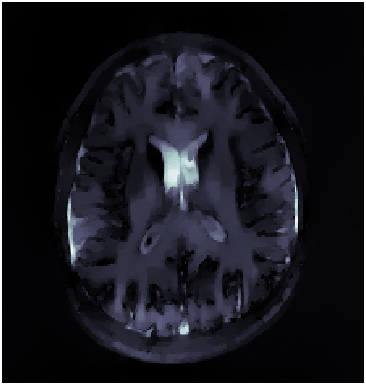}}
			%  \vspace{1.5cm}
			%. \centerline{(c) Result 4}\medskip 			% To add a manual subfigure caption
		\end{minipage}
		\hfill
		\begin{minipage}[b]{-6.4cm}
			\centering
			\centerline{\includegraphics[scale=0.55]{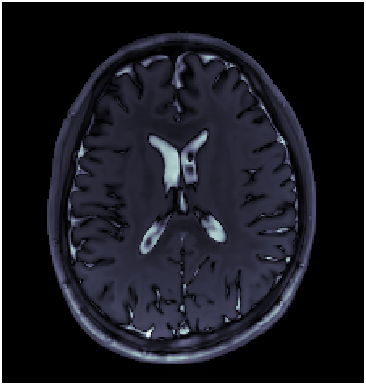}}
			%  \vspace{1.5cm}
			%. \centerline{(c) Result 4}\medskip 			% To add a manual subfigure caption
		\end{minipage}
		\\ % ----------------------------------------------------
		\begin{minipage}[b]{6.5cm}
			\centering
			\vspace{1.0mm}
			\centerline{\includegraphics[scale=0.55]{./figures/tsmis/ground_truth/tsmi_ground_truth_ch4_colorbar_horizontal_south.png}}
			\vspace{4.8mm}
			\centerline{(a) GT: Ch4}\medskip 		% To add a manual subfigure caption
		\end{minipage}
		\hfill
		\begin{minipage}[b]{-6.525cm}
			\centering
			\vspace{1.0mm}
			\centerline{\includegraphics[scale=0.55]{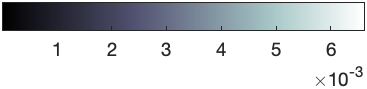}}
			\vspace{2.0mm}
			\centerline{(b) SVD-MRF: Ch4}\medskip 		% To add a manual subfigure caption
		\end{minipage}
		\hfill
		\begin{minipage}[b]{-6.45cm}
			\centering
			\vspace{1.0mm}
			\centerline{\includegraphics[scale=0.55]{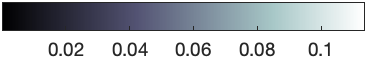}}
			\vspace{4.8mm}
			\centerline{(c) LRTV: Ch4}\medskip 		% To add a manual subfigure caption
		\end{minipage}
		\hfill
		\begin{minipage}[b]{-6.235cm}
			\centering
			\vspace{1.0mm}
			\centerline{\includegraphics[scale=0.55]{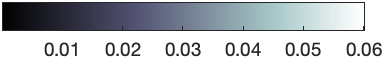}}
			\vspace{4.8mm}
			\centerline{(d) PnP-ADMM: Ch4}\medskip 		% To add a manual subfigure caption
		\end{minipage}
		%
		% ------------------------------------------------ Ch 4 --------------------------------------------------------
		%
		\\
		\vspace{3.5mm} %1.5, 2.5mm
		%
		% ------------------------------------------------ Ch 5 --------------------------------------------------------
		%
		\begin{minipage}[b]{6.4cm}
			\centering
			\centerline{\includegraphics[scale=0.55]{./figures/tsmis/ground_truth/tsmi_ground_truth_ch5.png}}
			%\vspace{-0.6mm}
			%  \centerline{(b) Results 3}\medskip 		% To add a manual subfigure caption
		\end{minipage}
		\hfill
		\begin{minipage}[b]{-6.4cm}
			\centering
			\centerline{\includegraphics[scale=0.55]{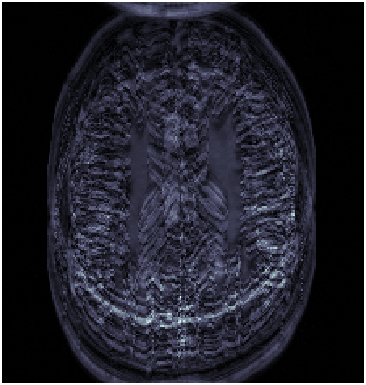}}
			%  \vspace{1.5cm}
			%  \centerline{(b) Results 3}\medskip 		% To add a manual subfigure caption
		\end{minipage}
		\hfill
		\begin{minipage}[b]{-6.4cm}
			\centering
			\centerline{\includegraphics[scale=0.55]{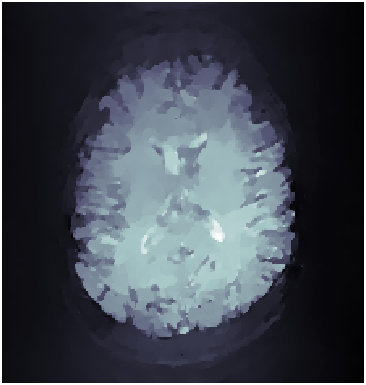}}
			%  \vspace{1.5cm}
			%. \centerline{(c) Result 4}\medskip 			% To add a manual subfigure caption
		\end{minipage}
		\hfill
		\begin{minipage}[b]{-6.4cm}
			\centering
			\centerline{\includegraphics[scale=0.55]{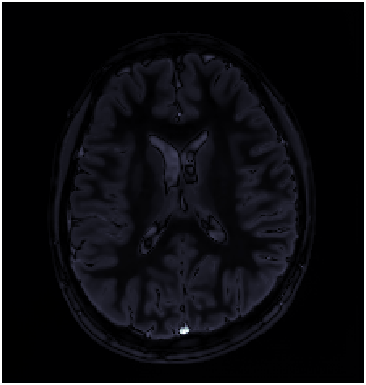}}
			%  \vspace{1.5cm}
			%. \centerline{(c) Result 4}\medskip 			% To add a manual subfigure caption
		\end{minipage}
		\\ % ----------------------------------------------------
		\begin{minipage}[b]{6.36cm}
			\centering
			\vspace{1.0mm}
			\centerline{\includegraphics[scale=0.55]{./figures/tsmis/ground_truth/tsmi_ground_truth_ch5_colorbar_horizontal_south.png}}
			\vspace{4.8mm}
			\centerline{(e) GT: Ch5}\medskip 		% To add a manual subfigure caption
		\end{minipage}
		\hfill
		\begin{minipage}[b]{-6.34cm}
			\centering
			\vspace{1.0mm}
			\centerline{\includegraphics[scale=0.55]{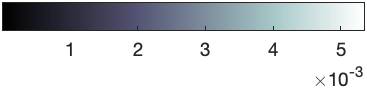}}
			\vspace{2.0mm}
			\centerline{(f) SVD-MRF: Ch5}\medskip 		% To add a manual subfigure caption
		\end{minipage}
		\hfill
		\begin{minipage}[b]{-6.24cm}
			\centering
			\vspace{1.0mm}
			\centerline{\includegraphics[scale=0.55]{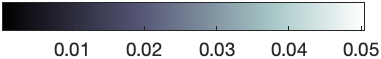}}
			\vspace{4.65mm}
			\centerline{(g) LRTV: Ch5}\medskip 		% To add a manual subfigure caption
		\end{minipage}
		\hfill
		\begin{minipage}[b]{-6.4cm}
			\centering
			\vspace{1.0mm}
			\centerline{\includegraphics[scale=0.55]{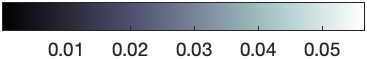}}
			\vspace{4.8mm}
			\centerline{(h) PnP-ADMM: Ch5}\medskip 		% To add a manual subfigure caption
		\end{minipage}
		%
		% ------------------------------------------------ Ch 5 --------------------------------------------------------
		%
		\\
		\vspace{3.5mm} %1.5mm, 2.5mm
		%
		% ------------------------------------------------ Ch 6 --------------------------------------------------------
		%
		\begin{minipage}[b]{6.4cm}
			\centering
			\centerline{\includegraphics[scale=0.55]{./figures/tsmis/ground_truth/tsmi_ground_truth_ch6.png}}
			%\vspace{-0.6mm}
			%  \centerline{(b) Results 3}\medskip 		% To add a manual subfigure caption
		\end{minipage}
		\hfill
		\begin{minipage}[b]{-6.4cm}
			\centering
			\centerline{\includegraphics[scale=0.55]{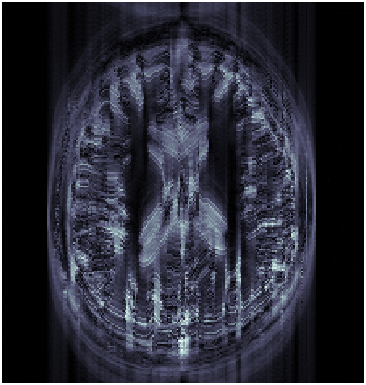}}
			%  \vspace{1.5cm}
			%  \centerline{(b) Results 3}\medskip 		% To add a manual subfigure caption
		\end{minipage}
		\hfill
		\begin{minipage}[b]{-6.4cm}
			\centering
			\centerline{\includegraphics[scale=0.55]{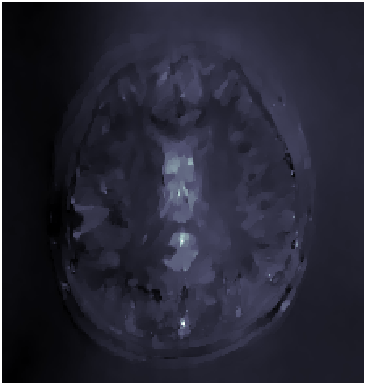}}
			%  \vspace{1.5cm}
			%. \centerline{(c) Result 4}\medskip 			% To add a manual subfigure caption
		\end{minipage}
		\hfill
		\begin{minipage}[b]{-6.4cm}
			\centering
			\centerline{\includegraphics[scale=0.55]{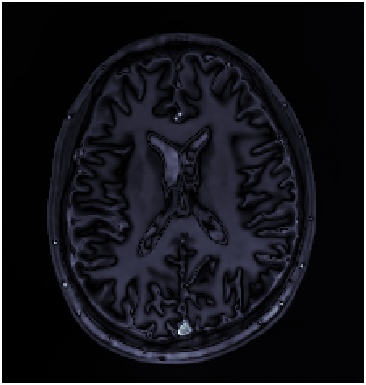}}
			%  \vspace{1.5cm}
			%. \centerline{(c) Result 4}\medskip 			% To add a manual subfigure caption
		\end{minipage}
		\\ % ----------------------------------------------------
		\begin{minipage}[b]{6.36cm}
			\centering
			\vspace{1.0mm}
			\centerline{\includegraphics[scale=0.55]{./figures/tsmis/ground_truth/tsmi_ground_truth_ch6_colorbar_horizontal_south.png}}
			\vspace{4.8mm}
			\centerline{(i) GT: Ch6}\medskip 		% To add a manual subfigure caption
		\end{minipage}
		\hfill
		\begin{minipage}[b]{-6.34cm}
			\centering
			\vspace{1.0mm}
			\centerline{\includegraphics[scale=0.55]{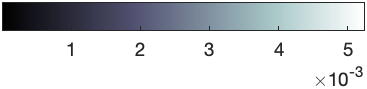}}
			\vspace{1.9mm}
			\centerline{(j) SVD-MRF: Ch6}\medskip 		% To add a manual subfigure caption
		\end{minipage}
		\hfill
		\begin{minipage}[b]{-6.36cm}
			\centering
			\vspace{1.0mm}
			\centerline{\includegraphics[scale=0.55]{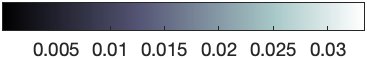}}
			\vspace{4.8mm}
			\centerline{(k) LRTV: Ch6}\medskip 		% To add a manual subfigure caption
		\end{minipage}
		\hfill
		\begin{minipage}[b]{-6.39cm}
			\centering
			\vspace{1.0mm}
			\centerline{\includegraphics[scale=0.55]{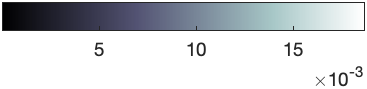}}
			\vspace{2.0mm}
			\centerline{(l) PnP-ADMM: Ch6}\medskip 		% To add a manual subfigure caption
		\end{minipage}
		%
		% ------------------------------------------------ Ch 6 --------------------------------------------------------
		%
		\\ 	% Need this here otherwise things will shift slightly
		\vspace{0.5cm}
	\caption{A visual comparison of the TSMIs obtained using the EPI Subsampling pattern for channels 4, 5 and 6, for slice 10.}
	\label{fig:supplementary_tsmi_fig_epi_2}
\end{center}

}]
% ---------- Supplementary Material - EPI TSMIs Fig - 2 (Camera Ready) ----------
% ---------- Supplementary Material - EPI TSMIs Fig - 3 (Camera Ready) ----------
\twocolumn[{
\renewcommand\twocolumn[1][]{#1}
% Reference: https://tex.stackexchange.com/questions/53966/how-to-place-a-banner-image-at-the-top-of-a-paper?rq=1
% Reference: https://tex.stackexchange.com/questions/53979/span-columns-with-a-center-environment/53984#53984
% Reference: https://tex.stackexchange.com/questions/55764/input-a-figure-between-title-and-body-in-twocolumn-form
\makeatletter 
\newcommand*\captiontype[1]{\def\@captype{#1}} 
\makeatother 

\begin{center}
	\captiontype{figure}
		\vspace{1.7cm} 	% 1.7cm, 1.75cm
		%
		% ------------------------------------------------ Ch 7 --------------------------------------------------------
		%
		\begin{minipage}[b]{6.4cm}
			\centering
			\centerline{\includegraphics[scale=0.55]{./figures/tsmis/ground_truth/tsmi_ground_truth_ch7.png}}
			%\vspace{-0.6mm}
			%  \centerline{(b) Results 3}\medskip 		% To add a manual subfigure caption
		\end{minipage}
		\hfill
		\begin{minipage}[b]{-6.4cm}
			\centering
			\centerline{\includegraphics[scale=0.55]{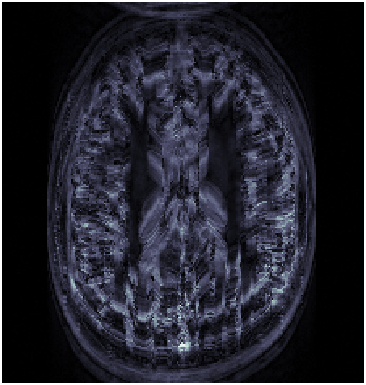}}
			%  \vspace{1.5cm}
			%  \centerline{(b) Results 3}\medskip 		% To add a manual subfigure caption
		\end{minipage}
		\hfill
		\begin{minipage}[b]{-6.4cm}
			\centering
			\centerline{\includegraphics[scale=0.55]{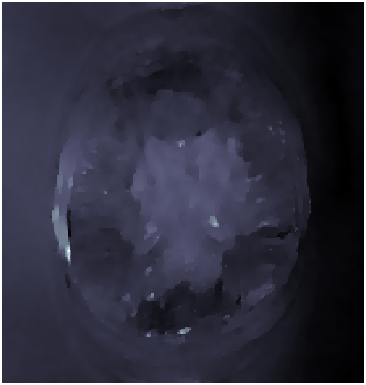}}
			%  \vspace{1.5cm}
			%. \centerline{(c) Result 4}\medskip 			% To add a manual subfigure caption
		\end{minipage}
		\hfill
		\begin{minipage}[b]{-6.4cm}
			\centering
			\centerline{\includegraphics[scale=0.55]{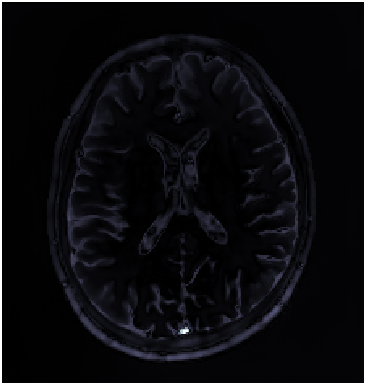}}
			%  \vspace{1.5cm}
			%. \centerline{(c) Result 4}\medskip 			% To add a manual subfigure caption
		\end{minipage}
		\\ % ----------------------------------------------------
		\begin{minipage}[b]{6.36cm}
			\centering
			\vspace{1.0mm}
			\centerline{\includegraphics[scale=0.55]{./figures/tsmis/ground_truth/tsmi_ground_truth_ch7_colorbar_horizontal_south.png}}
			\vspace{4.8mm}
			\centerline{(a) GT: Ch7}\medskip 		% To add a manual subfigure caption
		\end{minipage}
		\hfill
		\begin{minipage}[b]{-6.34cm}
			\centering
			\vspace{1.0mm}
			\centerline{\includegraphics[scale=0.55]{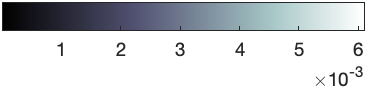}}
			\vspace{2.0mm}
			\centerline{(b) SVD-MRF: Ch7}\medskip 		% To add a manual subfigure caption
		\end{minipage}
		\hfill
		\begin{minipage}[b]{-6.25cm}
			\centering
			\vspace{1.0mm}
			\centerline{\includegraphics[scale=0.55]{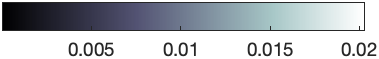}}
			\vspace{4.8mm}
			\centerline{(c) LRTV: Ch7}\medskip 		% To add a manual subfigure caption
		\end{minipage}
		\hfill
		\begin{minipage}[b]{-6.39cm}
			\centering
			\vspace{1.0mm}
			\centerline{\includegraphics[scale=0.55]{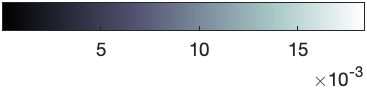}}
			\vspace{1.9mm}
			\centerline{(d) PnP-ADMM: Ch7}\medskip 		% To add a manual subfigure caption
		\end{minipage}
		%
		% ------------------------------------------------ Ch 7 --------------------------------------------------------
		%
		\\
		\vspace{3.5mm} %1.5, 2.5mm
		%
		% ------------------------------------------------ Ch 8 --------------------------------------------------------
		%
		\begin{minipage}[b]{6.4cm}
			\centering
			\centerline{\includegraphics[scale=0.55]{./figures/tsmis/ground_truth/tsmi_ground_truth_ch8.png}}
			%\vspace{-0.6mm}
			%  \centerline{(b) Results 3}\medskip 		% To add a manual subfigure caption
		\end{minipage}
		\hfill
		\begin{minipage}[b]{-6.4cm}
			\centering
			\centerline{\includegraphics[scale=0.55]{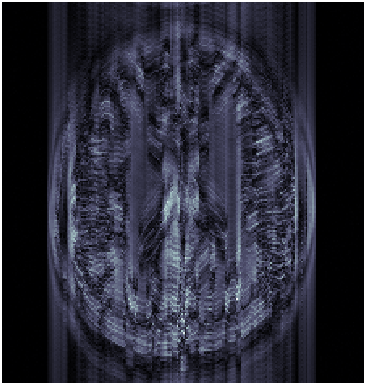}}
			%  \vspace{1.5cm}
			%  \centerline{(b) Results 3}\medskip 		% To add a manual subfigure caption
		\end{minipage}
		\hfill
		\begin{minipage}[b]{-6.4cm}
			\centering
			\centerline{\includegraphics[scale=0.55]{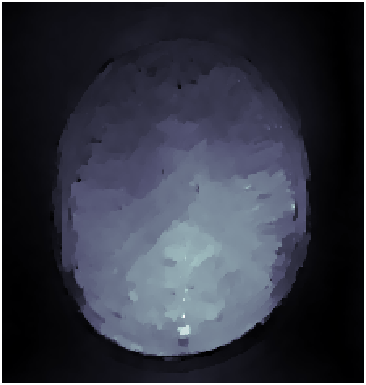}}
			%  \vspace{1.5cm}
			%. \centerline{(c) Result 4}\medskip 			% To add a manual subfigure caption
		\end{minipage}
		\hfill
		\begin{minipage}[b]{-6.4cm}
			\centering
			\centerline{\includegraphics[scale=0.55]{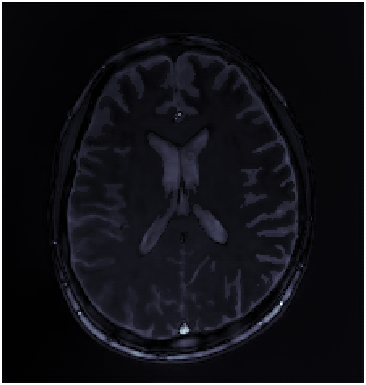}}
			%  \vspace{1.5cm}
			%. \centerline{(c) Result 4}\medskip 			% To add a manual subfigure caption
		\end{minipage}
		\\ % ----------------------------------------------------
		\begin{minipage}[b]{6.36cm}
			\centering
			\vspace{1.0mm}
			\centerline{\includegraphics[scale=0.55]{./figures/tsmis/ground_truth/tsmi_ground_truth_ch8_colorbar_horizontal_south.png}}
			\vspace{2.0mm}
			\centerline{(e) GT: Ch8}\medskip 		% To add a manual subfigure caption
		\end{minipage}
		\hfill
		\begin{minipage}[b]{-6.34cm}
			\centering
			\vspace{1.0mm}
			\centerline{\includegraphics[scale=0.55]{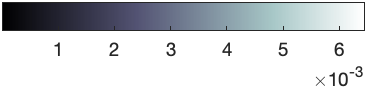}}
			\vspace{2.0mm}
			\centerline{(f) SVD-MRF: Ch8}\medskip 		% To add a manual subfigure caption
		\end{minipage}
		\hfill
		\begin{minipage}[b]{-6.26cm}
			\centering
			\vspace{1.0mm}
			\centerline{\includegraphics[scale=0.55]{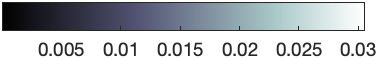}}
			\vspace{4.8mm}
			\centerline{(g) LRTV: Ch8}\medskip 		% To add a manual subfigure caption
		\end{minipage}
		\hfill
		\begin{minipage}[b]{-6.39cm}
			\centering
			\vspace{1.0mm}
			\centerline{\includegraphics[scale=0.55]{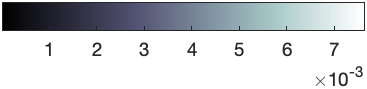}}
			\vspace{2.0mm}
			\centerline{(h) PnP-ADMM: Ch8}\medskip 		% To add a manual subfigure caption
		\end{minipage}
		%
		% ------------------------------------------------ Ch 8 --------------------------------------------------------
		%
		\\
		\vspace{3.5mm} %1.5, 2.5mm
		%
		% ------------------------------------------------ Ch 9 --------------------------------------------------------
		%
		\begin{minipage}[b]{6.4cm}
			\centering
			\centerline{\includegraphics[scale=0.55]{./figures/tsmis/ground_truth/tsmi_ground_truth_ch9.png}}
			%\vspace{-0.6mm}
			%  \centerline{(b) Results 3}\medskip 		% To add a manual subfigure caption
		\end{minipage}
		\hfill
		\begin{minipage}[b]{-6.4cm}
			\centering
			\centerline{\includegraphics[scale=0.55]{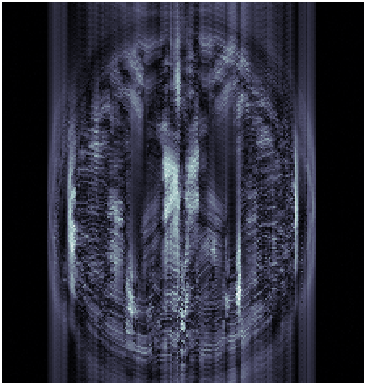}}
			%  \vspace{1.5cm}
			%  \centerline{(b) Results 3}\medskip 		% To add a manual subfigure caption
		\end{minipage}
		\hfill
		\begin{minipage}[b]{-6.4cm}
			\centering
			\centerline{\includegraphics[scale=0.55]{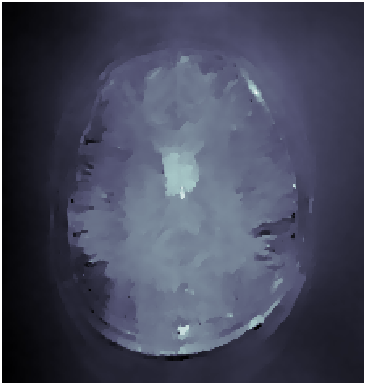}}
			%  \vspace{1.5cm}
			%. \centerline{(c) Result 4}\medskip 			% To add a manual subfigure caption
		\end{minipage}
		\hfill
		\begin{minipage}[b]{-6.4cm}
			\centering
			\centerline{\includegraphics[scale=0.55]{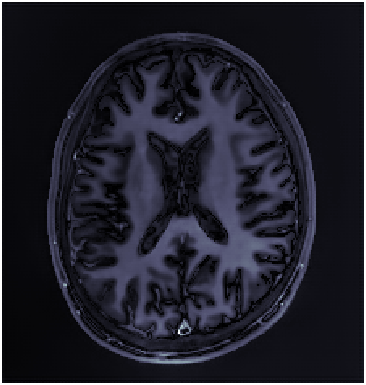}}
			%  \vspace{1.5cm}
			%. \centerline{(c) Result 4}\medskip 			% To add a manual subfigure caption
		\end{minipage}
		\\ % ----------------------------------------------------
		\begin{minipage}[b]{6.36cm}
			\centering
			\vspace{1.0mm}
			\centerline{\includegraphics[scale=0.55]{./figures/tsmis/ground_truth/tsmi_ground_truth_ch9_colorbar_horizontal_south.png}}
			\vspace{2.0mm}
			\centerline{(i) GT: Ch9}\medskip 		% To add a manual subfigure caption
		\end{minipage}
		\hfill
		\begin{minipage}[b]{-6.34cm}
			\centering
			\vspace{1.0mm}
			\centerline{\includegraphics[scale=0.55]{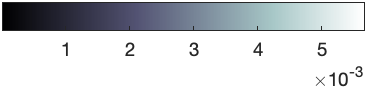}}
			\vspace{2.0mm}
			\centerline{(j) SVD-MRF: Ch9}\medskip 		% To add a manual subfigure caption
		\end{minipage}
		\hfill
		\begin{minipage}[b]{-6.37cm}
			\centering
			\vspace{1.0mm}
			\centerline{\includegraphics[scale=0.55]{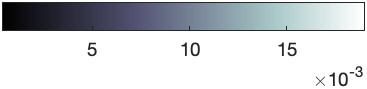}}
			\vspace{1.9mm}
			\centerline{(k) LRTV: Ch9}\medskip 		% To add a manual subfigure caption
		\end{minipage}
		\hfill
		\begin{minipage}[b]{-6.39cm}
			\centering
			\vspace{1.0mm}
			\centerline{\includegraphics[scale=0.55]{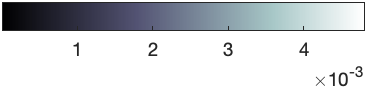}}
			\vspace{2.0mm}
			\centerline{(l) PnP-ADMM: Ch9}\medskip 		% To add a manual subfigure caption
		\end{minipage}
		%
		% ------------------------------------------------ Ch 9 --------------------------------------------------------
		%
		\\ 	% Need this here otherwise things will shift slightly
		\vspace{0.5cm}
	\caption{A visual comparison of the TSMIs obtained using the EPI Subsampling pattern for channels 7, 8 and 9, for slice 10.}
	\label{fig:supplementary_tsmi_fig_epi_3}
\end{center}

}]
% ---------- Supplementary Material - EPI TSMIs Fig - 3 (Camera Ready) ----------
% ---------- Supplementary Material - EPI TSMIs Fig - 4 (Camera Ready) ----------
\twocolumn[{
\renewcommand\twocolumn[1][]{#1}
% Reference: https://tex.stackexchange.com/questions/53966/how-to-place-a-banner-image-at-the-top-of-a-paper?rq=1
% Reference: https://tex.stackexchange.com/questions/53979/span-columns-with-a-center-environment/53984#53984
% Reference: https://tex.stackexchange.com/questions/55764/input-a-figure-between-title-and-body-in-twocolumn-form
\makeatletter 
\newcommand*\captiontype[1]{\def\@captype{#1}} 
\makeatother 

\begin{center}
	\captiontype{figure}
		\vspace{6.5cm} 	% 6.5cm
		%
		% ------------------------------------------------ Ch 10 --------------------------------------------------------
		%
		\begin{minipage}[b]{6.4cm}
			\centering
			\centerline{\includegraphics[scale=0.55]{./figures/tsmis/ground_truth/tsmi_ground_truth_ch10.png}}
			%\vspace{-0.6mm}
			%  \centerline{(b) Results 3}\medskip 		% To add a manual subfigure caption
		\end{minipage}
		\hfill
		\begin{minipage}[b]{-6.4cm}
			\centering
			\centerline{\includegraphics[scale=0.55]{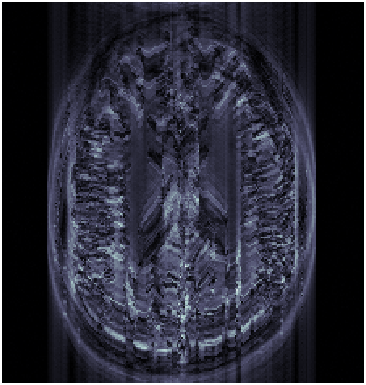}}
			%  \vspace{1.5cm}
			%  \centerline{(b) Results 3}\medskip 		% To add a manual subfigure caption
		\end{minipage}
		\hfill
		\begin{minipage}[b]{-6.4cm}
			\centering
			\centerline{\includegraphics[scale=0.55]{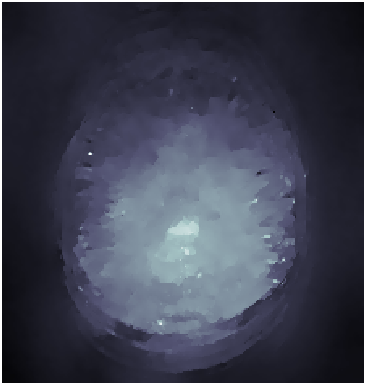}}
			%  \vspace{1.5cm}
			%. \centerline{(c) Result 4}\medskip 			% To add a manual subfigure caption
		\end{minipage}
		\hfill
		\begin{minipage}[b]{-6.4cm}
			\centering
			\centerline{\includegraphics[scale=0.55]{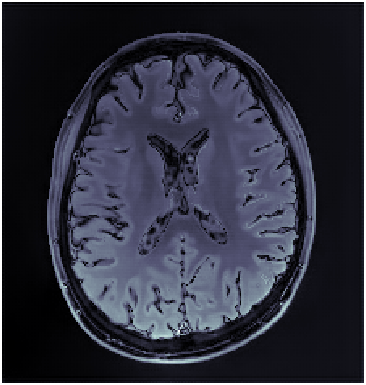}}
			%  \vspace{1.5cm}
			%. \centerline{(c) Result 4}\medskip 			% To add a manual subfigure caption
		\end{minipage}
		\\ % ----------------------------------------------------
		\begin{minipage}[b]{6.36cm}
			\centering
			\vspace{1.0mm}
			\centerline{\includegraphics[scale=0.55]{./figures/tsmis/ground_truth/tsmi_ground_truth_ch10_colorbar_horizontal_south.png}}
			\vspace{2.0mm}
			\centerline{(a) GT: Ch10}\medskip 		% To add a manual subfigure caption
		\end{minipage}
		\hfill
		\begin{minipage}[b]{-6.34cm}
			\centering
			\vspace{1.0mm}
			\centerline{\includegraphics[scale=0.55]{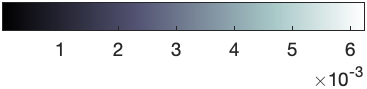}}
			\vspace{1.9mm}
			\centerline{(b) SVD-MRF: Ch10}\medskip 		% To add a manual subfigure caption
		\end{minipage}
		\hfill
		\begin{minipage}[b]{-6.35cm}
			\centering
			\vspace{1.0mm}
			\centerline{\includegraphics[scale=0.55]{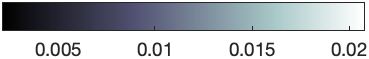}}
			\vspace{4.8mm}
			\centerline{(c) LRTV: Ch10}\medskip 		% To add a manual subfigure caption
		\end{minipage}
		\hfill
		\begin{minipage}[b]{-6.39cm}
			\centering
			\vspace{1.0mm}
			\centerline{\includegraphics[scale=0.55]{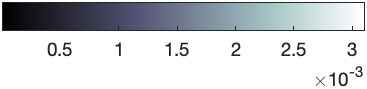}}
			\vspace{2.0mm}
			\centerline{(d) PnP-ADMM: Ch10}\medskip 		% To add a manual subfigure caption
		\end{minipage}
		%
		% ------------------------------------------------ Ch 10 --------------------------------------------------------
		%
		\\ 	% Need this here otherwise things will shift slightly
		\vspace{0.5cm}
	\caption{A visual comparison of the TSMIs obtained using the EPI Subsampling pattern for channel 10, for slice 10.}
	\label{fig:supplementary_tsmi_fig_epi_4}
\end{center}

}]
% ---------- Supplementary Material - EPI TSMIs Fig - 4 (Camera Ready) ----------
%
% ---------------------------------------------------------- EPI ----------------------------------------------------------
%